\documentclass[a4paper,11pt]{article}
\usepackage{bbm}
\pdfoutput=1
\usepackage{jheppub}
\usepackage{epsfig}
\usepackage{amssymb}
\usepackage{amsmath}
\usepackage[active]{srcltx}
\usepackage{slashed}

\newcommand{\bea}{\begin{eqnarray}}
\newcommand{\eea}{\end{eqnarray}}
\newcommand{\bean}{\begin{eqnarray*}}
\newcommand{\eean}{\end{eqnarray*}}

\def\braket#1{\left\langle #1 \right\rangle}

\def\gb #1{ \left\langle #1 \right]}

\def\Tr{\mathop{\rm Tr}}

\def\vev{\braket}

\def\bvev#1{\left[ #1 \right]}

\def\Label#1{\label{#1}%
  \smash{\hbox to0pt{\raise1ex\hbox{\tiny[#1]}\hss}}}

\def\Spaa{\vev}
\def\Spbb{\bvev}
\def\Spab{\gb}

\def\SYM{{\tiny\mbox{SYM}}}
\def\MHV{{\tiny \mbox{MHV}}}
\def\spaa #1{\langle #1\rangle}
\def\spbb #1{[#1]}
\def\spab #1{\langle #1]}

\def\I{\tiny \mbox{I}}
\def\II{\tiny \mbox{II}}
\def\III{\tiny \mbox{III}}

\title{Form Factor and Boundary Contribution of Amplitude}
\author[a]{Rijun Huang,}
\author[a]{Qingjun Jin\footnote{The correspondence author.}}
\author[a,b]{and Bo Feng\footnote{The
unusual ordering of authors is just to let authors get proper
recognition of contributions under outdated practice in China. }}

\affiliation[a]{Zhejiang Institute of Modern Physics, Department of Physics, Zhejiang University, Hangzhou, 310027, P. R. China}
\affiliation[a]{Center of Mathematical Science, Zhejiang University, Hangzhou, 310027, P. R. China}

\emailAdd{huang@nbi.dk}
\emailAdd{qingjun@zju.edu.cn}
\emailAdd{fengbo@zju.edu.cn}

\abstract{The boundary contribution of an amplitude in the BCFW recursion relation
can be considered as a form factor involving boundary operator and unshifted particles.
At the tree-level, we show that by suitable construction of Lagrangian, one can relate
the leading order term of  boundary operators to some composite operators of $\mathcal{N}=4$ super-Yang-Mills theory,
then the computation of form factors is translated to the computation of amplitudes. We compute
the form factors of these composite operators through the computation of corresponding double
trace amplitudes.}

\begin{document}
%%%%%%%%%%%%%%%%%%%%%%%%%%%%%%%%%%%%%%%%%%%%%%%%%%%%%%%%%%%%%%%%%%
\maketitle \flushbottom

%%%%%%%%%%%%%%%%%%%
\section{Introduction}
%%%%%%%%%%%%%%%%%%%%%

The ongoing research on the on-shell techniques has gone beyond its
primal scattering amplitude domain, to the computation of form
factor in recent years. The form factor, sometimes stated as a
bridge linking on-shell amplitude and off-shell correlation
function, is a quantity containing both on-shell states(ingredients
for amplitudes) and gauge invariant operators(ingredients for
correlation functions). Its computation can be traced back to the
pioneering paper\cite{vanNeerven:1985ja} nearly 30 years ago, where
the Sudakov form factor of the bilinear scalar operator
$\Tr(\phi^2)$ is investigated up to two loops. At present, many
revolutionary insights originally designed for the computation of
amplitudes\footnote{See reviews, e.g.,
\cite{Bern:2007dw,Elvang:2013cua,Henn:2014yza}.}, such as MHV vertex
expansion\cite{Cachazo:2004kj}, BCFW recursion
relation\cite{Britto:2004ap,Britto:2005fq}, color-kinematic
duality\cite{Bern:2008qj,Bern:2010ue}, unitarity cut
\cite{Bern:1994zx,Bern:1994cg} method(and its generalization to
$D$-dimension \cite{Anastasiou:2006jv,Anastasiou:2006gt}),
generalized unitarity\cite{Britto:2004nc,Britto:2005ha}, etc., have
played their new roles in evaluating form factors.

These progresses are achieved in various papers. In paper
\cite{Brandhuber:2010ad}, the BCFW recursion relation appears for
the first time in the recursive computation of tree-level form
factor, mainly for the bilinear scalar operator. As a consequence,
the solution of recursion relation for split helicity form factor is
conquered\cite{Brandhuber:2011tv}. Intensive discussion on the
recursion relation of form factor is provided later in
\cite{Bork:2014eqa}. A generalization to the form factor of full
stress tensor multiplet is discussed in \cite{Brandhuber:2011tv} and
\cite{Bork:2010wf}, where in the former one, supersymmetric version
of BCFW recursion relation is pointed out to be applicable to super
form factor. Shortly after, the color-kinematic duality is
implemented in the context of form factor\cite{Boels:2012ew}, both
at tree and loop-level, to generate the integrand of form factor.
Most recently, the elegant formulation of amplitudes based on
Grassmannian prescription\cite{ArkaniHamed:2012nw} is also extended
to tree-level form factors\cite{Frassek:2015rka}. At loop-level, the
form factor is generally computed by unitarity cut method. The
generic Maximal-Helicity-Violating(MHV) super form factor as well as
some Next-MHV(NMHV) form factor at one-loop are computed in
\cite{Brandhuber:2011tv,Bork:2011cj,Bork:2012tt,Engelund:2012re}
with compact results. The Sudakov form factor is computed to three
loops in \cite{Gehrmann:2006wg,Baikov:2009bg,Gehrmann:2011xn}. The
three-point two-loop form factor of half-BPS operator is achieved in
\cite{Brandhuber:2012vm}, and the general $n$-point form factor as
well as the remainder functions in \cite{Brandhuber:2014ica}. The
scalar operator with arbitrary number of scalars is discussed in
\cite{Bork:2010wf,Penante:2014sza,Brandhuber:2014ica}. Beyond the
half-BPS operators, form factors of non-protected operators, such as
dilatation operator\cite{Wilhelm:2014qua}, Konishi
operator\cite{Nandan:2014oga}, operators in the $SU(2)$
sectors\cite{Loebbert:2015ova}, are also under investigation.
Furthermore, the soft theorems for the form factor of half-BPS and
Konishi operators are studied at tree and one-loop
level\cite{Bork:2015fla}, showing similarity to amplitude case.
Carrying on the integrand result of \cite{Boels:2012ew}, the master
integrals for four-loop Sudakov form factor is determined in
\cite{Boels:2015yna}. An alternative discussion on the master
integrals of form factor in massless QCD can be found in
\cite{vonManteuffel:2015gxa}. Similar unitarity based studies on
Sudakov form factor of three-dimensional ABJM theories are also
explored\cite{Brandhuber:2013gda,Young:2013hda,Bianchi:2013pfa}.

The above mentioned achievements encode the belief that the
state-of-art on-shell techniques of amplitude would also be
applicable to form factor. Recently, the advances in the computation
of boundary contribution have revealed another connection between
form factor and amplitude. When talking about the BCFW recursion
relation of amplitude, the boundary contribution is generally
assumed to be absent. However this assumption is not always true,
for example, it fails in the theories involving only scalars and
fermions or under the "bad" momentum deformation. Many solutions
have been proposed(by auxiliary
fields\cite{Benincasa:2007xk,Boels:2010mj}, analyzing Feynman
diagrams\cite{Feng:2009ei,Feng:2010ku,Feng:2011twa}, studying the
zeros\cite{Benincasa:2011kn,Benincasa:2011pg,Feng:2011jxa}, the
factorization limits\cite{Zhou:2014yaa}, or using other
deformation\cite{Cheung:2015cba,Cheung:2015ota,Luo:2015tat}) to deal
with the boundary contribution in various situations. Most recently,
a new multi-step BCFW recursion relation
algorithm\cite{Feng:2014pia,Jin:2014qya,Feng:2015qna} is proposed to
detect the boundary contribution through certain poles step by step.
Especially in paper \cite{Jin:2014qya}, it is pointed out that the
boundary contribution possesses similar BCFW recursion relation as
amplitudes, and it can be computed recursively from the lower-point
boundary contribution. Based on this idea, later in paper
\cite{Jin:2015pua}, the boundary contribution is further interpreted
as form factor of certain composite operator named {\sl boundary
operator}, while the boundary operator can be extracted from the
operator product expansion(OPE) of deformed fields.

The idea of boundary operator motives us to connect the computation
of form factor to the boundary contribution of amplitudes. Since a
given boundary contribution of amplitude can be identified as a form
factor of certain boundary operator, we can also interpret a given
form factor as the boundary contribution of certain amplitude. In
paper \cite{Jin:2015pua}, the authors showed how to construct the
boundary operator starting from a known Lagrangian. We can reverse
the logic and ask the question: for a given operator, how can we
construct a Lagrangian whose boundary operator under certain
momentum deformation is exactly the operator of request? In this
paper, we try to answer this question by constructing the Lagrangian
for a class of so called composite operators. Once the Lagrangian is
ready, we can compute the corresponding amplitude, take appropriate
momentum shifting and extract the boundary contribution, which is
identical(or proportional) to the form factor of that operator. By
this way, the computation of form factor can be considered as a
problem of computing the amplitude of certain theory.

This paper is structured as follows. In \S \ref{secReview}, we
briefly review the BCFW recursion relation and boundary operator. We
also list the composite operators of interest, and illustrate how to
construct the Lagrangian that generates the boundary operators of
request. In \S \ref{secSudakov}, using Sudakov form factor as
example, we explain how to compute the form factor through computing
the boundary contribution of amplitude, and demonstrate the
computation by recursion relation of form factor, amplitude and
boundary contribution. We show that these three ways of
understanding lead to the same result. In \S \ref{secComposite}, we
compute the form factors of composite operators by constructing
corresponding Lagrangian and working out the amplitude of double
trace structure. Conclusion and discussion can be found in \S
\ref{secConclusion}, while in the appendix, the construction of
boundary operator starting from Lagrangian is briefly reviewed for
reader's convenience, and the discussion on large $z$ behavior is
presented.

%%%%%%%%%%%%%%%%%%
\section{From boundary contribution to form factor}
\label{secReview}
%%%%%%%%%%%%%%%%%%

The BCFW recursion relation \cite{Britto:2004ap,Britto:2005fq}
provides a new way of studying scattering amplitude in S-matrix
framework. Using suitable momentum shifting, for example,
\bea \widehat{p}_i=p_i-z
q~~,~~\widehat{p}_j=p_j+zq~~\mbox{while}~~q^2=p_i\cdot q=p_j\cdot
q=0~,~~~\label{bcfw-shifting}\eea
one can treat the amplitude as an analytic function $A(z)$ of single
complex variable, with poles in finite locations and possible
non-vanishing terms in boundary, while the physical amplitude sits
at $z=0$ point. Assuming that under certain momentum shifting,
$A(z)$ has no boundary contribution in the contour integration
${1\over 2\pi i}\oint {dz\over z}A(z)$, i.e., $A(z)\to 0$ when $z\to
\infty$, then the physical amplitude $A(z=0)$ can be purely
determined by the residues of $A(z)$ at finite poles. However, if
$A(z)$ does not vanish around the infinity, for example when taking
a "bad" momentum shifting or in theories such as $\lambda\phi^4$,
the boundary contribution would also appear as a part of physical
amplitude. Most people would try to avoid dealing with such theories
as well as the "bad" momentum shifting, since the evaluation of
boundary contribution is much more complicated than taking the
residues of $A(z)$.

Although it is usually unfavored during the direct computation of
amplitude, authors in paper \cite{Jin:2015pua} found that the
boundary contribution is in fact {\sl a form factor involving
boundary operator and unshifted particles},
\bea B^{\spab{1|2}}=\spaa{\Phi(p_3)\cdots
\Phi(p_{n})|\mathcal{O}^{\spab{1|2}}(0)|0}~,~~~\label{boundaryFrom}\eea
where $\Phi(p_i)$ denotes arbitrary on-shell fields, and momenta of
$\Phi(p_1),\Phi(p_2)$ have been shifted according to
eqn.(\ref{bcfw-shifting}). The momentum $q$ carried by the boundary
operator is $q=-p_1-p_2=\sum_{i=3}^n p_i$. Eqn. (\ref{boundaryFrom})
is identical to a $(n-2)$-point form factor generated by operator
$\mathcal{O}^{\spab{1|2}}$ with off-shell momentum $q^2\neq 0$. The
observation (\ref{boundaryFrom}) provides a new way of computing
form factor,
\begin{enumerate}
  \item Construct the Lagrangian, and compute the corresponding amplitude,
  \item Take the appropriate momentum shifting, and pick up the boundary
contribution,
  \item Read out the form factor from boundary contribution after
  considering LSZ reduction.
\end{enumerate}

In paper \cite{Jin:2015pua}, the authors illustrated how to work out
the boundary operator $\mathcal{O}^{\spab{\Phi_i|\Phi_j}}$ from
Lagrangian of a given theory under momentum shifting of two selected
external fields. Starting from a Lagrangian, one can eventually
obtain a boundary operator. For example, a real massless scalar
theory with $\phi^m$ interaction
\bea L=-{1\over 2} (\partial \phi)^2+{\kappa\over m!}\phi^m~,~~~\eea
under momentum shifting of two scalars(say $\phi_1$ and $\phi_2$)
will produce a boundary operator
\bea \mathcal{O}^{\spab{\phi_1|\phi_2}}={\kappa\over
(m-2)!}\phi^{m-2}~.~~~\eea
Hence the boundary contribution of a $n$-point amplitude
$A_n(\phi_1,\ldots, \phi_n)$ in this $\kappa\phi^m$ theory under
$\spab{\phi_1|\phi_2}$-shifting is identical to the $(n-2)$-point
form factor
\bea
\mathcal{F}_{\mathcal{O}^{\spab{\phi_1|\phi_2}},n-2}(\phi_3,\ldots,\phi_{n};q)\equiv{\kappa\over
(m-2)!}\spaa{\phi_3\cdots\phi_{n}|\phi^{m-2}(0)|0}~.~~~\eea
However, this form factor is not quite interesting. We are
interested in certain kind of operators, such as bilinear half-BPS
scalar operator $\Tr(\phi^{AB}\phi^{AB})$ or chiral stress-tensor
operator $\Tr(W^{++}W^{++})$ in $\mathcal{N}=4$
super-Yang-Mills(SYM) theory, where $W^{++}$ is a particular
projection of the chiral vector multiplet superfield
$W^{AB}(x,\theta)$ in SYM. What we want to do is to compute the form
factor for a given operator, but not the operators generated from
arbitrary Lagrangian. More explicitly, if we want to compute the
form factor of operator $\mathcal{O}$, we should first construct a
Lagrangian whose boundary operator is identical(or proportional) to
$\mathcal{O}$. With such Lagrangian in hand, we can then compute the
corresponding amplitude, take the momentum shifting and pick up the
boundary contribution. So the problem is how to construct the
corresponding Lagrangian.

%%%%%%%%%%%%%%%%%%%
\subsection{The operators of interest}
\label{secOperator}
%%%%%%%%%%%%%%%%

It is obvious that the construction of Lagrangian depends on the
operators we want to produce. In this paper, we will study the so
called gauge-invariant local composite operators, which are built as
traces of product of gauge-covariant fields at a common spacetime
point. These fields are taken to be the component fields of
$\mathcal{N}=4$ superfield $\Phi^{\mathcal{N}=4}$
\cite{Nair:1988bq}, given by six real scalars $\phi^I,I=1,\ldots,
6$(or 3 complex scalars $\phi^{AB}$), four fermions
$\psi^A_{\alpha}=\epsilon^{ABCD}\psi_{BCD\alpha}$, four
anti-fermions $\bar{\psi}_{A\dot{\alpha}}$ and the field strength
$F_{\mu\nu}$, where $\alpha,\beta,\dot{\alpha},\dot{\beta}=1,2$ are
spinor indices, $A,B,C,D=1,2,3,4$ are $SU(4)$ R-symmetric indices,
and $\mu,\nu=0,1,2,3$ are spacetime indices. The field strength can
be further split into self-dual and anti-self-dual parts
$F_{\alpha\beta},\bar{F}_{\dot{\alpha}\dot{\beta}}$:
\bea
F_{\alpha\beta\dot{\alpha}\dot{\beta}}=F_{\mu\nu}(\sigma^{\mu})_{\alpha\dot{\alpha}}(\sigma^{\nu})_{\beta\dot{\beta}}
=\sqrt{2}\epsilon_{\dot{\alpha}\dot{\beta}}F_{\alpha\beta}+\sqrt{2}\epsilon_{\alpha\beta}\bar{F}_{\dot{\alpha}\dot{\beta}}~,~~~\eea
corresponding to positive gluon and negative gluon respectively.

The number of fields inside the trace is called the length of
operator, and the simplest non-trivial ones are the length two
operators. There is no limit on the length of operator, for example,
the bilinear half-BPS scalar operator $\Tr(\phi^I\phi^J)$ is length
two, while we could also have length $L$ scalar operator
$\Tr(\phi^{I_1}\cdots \phi^{I_L})$. The operators can also carry
spinor indices, such as
$\mathcal{O}^{\alpha\beta\dot{\alpha}\dot{\beta}}=\Tr(\psi^{A\alpha}\psi^{B\beta}\bar{F}^{\dot{\alpha}\dot{\beta}})$
in the $(1,1)$ representation under Lorentz group $SU(2)\times
SU(2)$.

We will mainly focus on the length two operators. These operators
can be classified by their spins and labeled by their
representations under $SU(2)\times SU(2)$ group. For spin-0
operators in $(0,0)$-representation, we have
\bea
&&\mathcal{O}^{[0]}_{\I}=\Tr(\phi^{I}\phi^{J})~~,~~\mathcal{O}^{[0]}_{\II}=\Tr(\psi^{A\alpha}\psi^{B}_{\alpha})~~,~~
\mathcal{O}^{[0]}_{\III}=\Tr(F^{\alpha\beta}F_{\alpha\beta})~,~~~\nonumber\\
&&~~~~~~~~~~~~~~~~~~~~~~~~~~\bar{\mathcal{O}}^{[0]}_{\II}=\Tr(\bar{\psi}^{\dot{\alpha}}_{A}\bar{\psi}_{B\dot{\alpha}})~~,~~
\bar{\mathcal{O}}^{[0]}_{\III}=\Tr(\bar{F}^{\dot{\alpha}\dot{\beta}}\bar{F}_{\dot{\alpha}\dot{\beta}})~.~~~\label{spin0-operator}\eea
For spin-${1\over 2}$ operators in $({1\over 2},0)$ or $(0,{1\over
2})$-representation, we have
\bea
&&\mathcal{O}^{[1/2]}_{\I}=\Tr(\phi^{I}\psi^{A\alpha})~~,~~\mathcal{O}^{[1/2]}_{\II}=\Tr(\psi^{A}_{\beta}F^{\beta\alpha})~,~~~\nonumber\\
&&\bar{\mathcal{O}}^{[1/2]}_{\I}=\Tr(\phi^{I}\bar{\psi}_{A}^{\dot{\alpha}})~~,~~\bar{\mathcal{O}}^{[1/2]}_{\II}=\Tr(\bar{\psi}_{A\dot{\beta}}\bar{F}^{\dot{\beta}\dot{\alpha}})~.~~~\label{spin12-operator}\eea
For spin-1 operators in $(1,0)$ or $(0,1)$-representation, we have
\bea
&&\mathcal{O}^{[1]}_{\I}=\Tr(\psi^{A\alpha}\psi^{B\beta}+\psi^{A\beta}\psi^{B\alpha})~~,~~\mathcal{O}^{[1]}_{\II}=\Tr(\phi^{I}
F^{\alpha\beta})~,~~~\nonumber\\
&&\bar{\mathcal{O}}^{[1]}_{\I}=\Tr(\bar{\psi}_A^{~~\dot{\alpha}}\bar{\psi}_B^{~~\dot{\beta}}+\bar{\psi}_A^{~~\dot{\beta}}\bar{\psi}_B^{~~\dot{\alpha}})~~,~~\bar{\mathcal{O}}^{[1]}_{\II}=\Tr(\phi^{I}
\bar{F}^{\dot{\alpha}\dot{\beta}})~,~~~\label{spin1-operator1}\eea
and in $({1\over 2},{1\over 2})$-representation,
\bea
\mathcal{O}^{[1]}_{\III}=\Tr(\psi^{A\alpha}\bar{\psi}_B^{\dot{\alpha}})~.~~~\label{spin1-operator2}\eea
For spin-${3\over 2}$ operators in $(1,{1\over 2})$ or $({1\over
2},1)$-representation, we have
\bea
\mathcal{O}^{[3/2]}_{\I}=\Tr(\bar{\psi}_A^{\dot{\alpha}}F^{\alpha\beta})~~,~~\bar{\mathcal{O}}^{[3/2]}_{\I}=\Tr({\psi}^{A{\alpha}}\bar{F}^{\dot{\alpha}\dot{\beta}})~.~~~\label{spin32-operator1}\eea
and in $({3\over 2},0)$ or $(0,{3\over 2})$-representation,
\bea
\mathcal{O}^{[3/2]}_{\II}=\Tr({\psi}^{A\gamma}{F}^{{\alpha}{\beta}})~~,~~\bar{\mathcal{O}}^{[3/2]}_{\II}=\Tr(\bar{\psi}_A^{\dot{\gamma}}\bar{F}^{\dot{\alpha}\dot{\beta}})~.~~~\label{spin32-operator2}\eea
For spin-2 operators in $(1,1)$-representation, we have
\bea
\mathcal{O}^{[2]}_{\I}=\Tr(F^{\alpha\beta}\bar{F}^{\dot{\alpha}\dot{\beta}})~.~~~\label{spin2-operator}\eea
For operators of the same class, we can apply similar procedure to
construct the Lagrangian. The operators with length larger than two
can be similarly written down, and classified according to their
spins and representations. For those whose spins are no larger than
2, we can apply the same procedure as is done for length two
operators. while if their spins are larger than 2, we need multiple
shifts.

Some of above operators are in fact a part of the chiral
stress-tensor multiplet operator in $\mathcal{N}=4$ SYM
\cite{Eden:2011yp,Eden:2011ku}, and their form factors are
components of $\mathcal{N}=4$ super form factor. However, we have
assumed that, all indices of these gauge-covariant fields are
general, so above operators are not limited to the chiral part, they
are quite general.

%%%%%%%%%%%%%%%%%%%%
\subsection{Constructing the Lagrangian}
%%%%%%%%%%%%%%%%%%%%%%

One important property shared by above operators is that they are
all traces of fields. Tree-level amplitudes of ordinary gauge theory
only possess single trace structure. From the shifting of two
external fields, one can not generate boundary operators with trace
structures, which can be seen in \cite{Jin:2015pua}. The solution is
to intentionally add a double trace term in the standard Lagrangian.
The added term should be gauge-invariant, and generate the
corresponding operator under selected momentum shifting.

For a given operator $\mathcal{O}$ of interest, let us add a double
trace term $\Delta L$ to the $\mathcal{N}=4$ Lagrangian $L_{\SYM}$,
\bea L_{\mathcal{O}}=L_{\SYM}+
\frac{\kappa}{N}\Tr(\Phi^{\alpha'_1}\Phi^{\alpha'_2})\mathcal{O}
+\frac{\bar{\kappa}}{N}\Tr(\Phi^{\dagger}_{\alpha'_1}\Phi^{\dagger}_{\alpha'_2})\bar{\mathcal{O}}~,~~~\label{lagrangianO}
\eea
where $SU(N)$ group is assumed, $\kappa,\bar{\kappa}$ are coupling
constants for the double trace interactions(which can be re-scaled
to fit the overall factor of final result) and $\Phi^{\alpha'}$,
$\Phi^{\dagger}_{\alpha'}$ denotes\footnote{The definition of $\Phi,
\Phi^{\dagger}$ can be found in (\ref{defPhi}), and remind that the
index here of $\Phi,\Phi^{\dagger}$ is not spinor index but the
index of their components, which specifies $\Phi$ to be scalar,
fermion or gluon.} any type of fields among
$\phi^I,\psi^{A\alpha},\bar{\psi}_A^{\dot{\alpha}},F^{\alpha\beta},\bar{F}^{\dot{\alpha}\dot{\beta}}$.
The spinor indices are not explicitly written down for $\Phi,
\Phi^{\dagger}$, however we note that they should be contracted with
the spinor indices of the operator, so that the added Lagrangian
terms are Lorentz invariant. We will show that at the large $N$
limit, momentum shifting of two fields in $\Delta L$ indeed
generates the boundary operator $\mathcal{O}$.

\def\full{{\tiny \mbox{full}}}
The tree-level amplitudes defined by Lagrangian $L_{\mathcal{O}}$
can have single trace pieces or multiple trace pieces. A full
$(n+2)$-point amplitude
$$A_{n+2}^{\full}(\Phi^{\alpha_1a_1},\ldots,
\Phi^{\alpha_na_n},\Phi^{\alpha_{n+1}a},\Phi^{\alpha_{n+2}b})$$ thus
can be decomposed into color-ordered partial amplitudes $A$ as
\bea
A_{n+2}^{\full}&=&A_{n+2}(1,2,\ldots,n+2)\Tr(t^{a_1}\cdots t^{a_n}t^at^b)+\cdots\label{aoriginal}\\
&&+\frac{1}{N}A_{k;n+2-k}(1,\ldots,k;k+1,\ldots ,
n+2)\Tr(t^{a_1}\cdots t^{a_{k}})\Tr(t^{a_{k+1}}\cdots
t^at^b)+\cdots\nonumber\eea
where $A_{n}$ denotes $n$-point single trace amplitude, $A_{k;n-k}$
denotes $n$-point double trace amplitude. We use $i$ to abbreviate
$\Phi_i$, and $\cdots$ stands for all possible permutation terms and
other higher order multiple trace pieces. Since the operator
$\mathcal{O}$ we want to generate is single trace, the terms with
higher multiple trace in $\cdots$ is then irrelevant for our
discussion, and also they can be ignored at large $N$. Now let us
contract the color indices $a,b$, which gives\footnote{Remind the
identity
$(t^a)^{~\bar{\jmath}_1}_{i_1}(t^a)^{~\bar{\jmath}_2}_{i_2}=\delta^{~\bar{\jmath}_2}_{i_1}\delta^{~\bar{\jmath}_1}_{i_2}-{1\over
N}\delta^{~\bar{\jmath}_1}_{i_1}\delta^{~\bar{\jmath}_2}_{i_2}$.}
\bea A_{n+2}^{\full}&=&{N^2-1\over
N}A_{n+2}(1,2,\ldots,n+2)\Tr(t^{a_1}\cdots t^{a_n})+\cdots\label{acontract}\\
&&+{N^2-1\over N^2}A_{k;n+2-k}(1,\ldots,k;k+1,\ldots,
n+2)\Tr(t^{a_1}\cdots t^{a_k})\Tr(t^{a_{k+1}}\cdots
t^{a_n})+\cdots\nonumber\eea
In this case, the $O(N)$ order terms in (\ref{acontract}) come from
two places, one is the single trace part in (\ref{aoriginal}) when
$t^a$ and $t^b$ are adjacent, the other is the double trace part in
(\ref{aoriginal}) whose color factor has the form $\Tr(\cdots )\Tr(
t^at^b)$. So when color indices $a,b$ are contracted, the leading
contribution of the full $(n+2)$-point amplitude is
\bea A_{n+2}^{\full}=N\Tr(t^{a_1}\cdots
t^{a_n})\mathcal{K}(1,2,\ldots,
n)+\mbox{possible~permutation}\{1,2,\ldots,n\}~,~~~\eea
where
\bea \mathcal{K}(1,\ldots, n)&\equiv&A_{n+2}(1,\ldots,
n,n+1,n+2)+A_{n+2}(1,\ldots, n,n+2,n+1)\nonumber\\
&&+A_{n;2}(1,\ldots, n;n+1,n+2)~.~~~\label{aconN}\eea
The first two terms in $\mathcal{K}$ are the same as the
corresponding color-ordered single trace amplitudes, since the other
double trace terms in the Lagrangian will not contribute to the
$O(N)$ order at tree-level. The third term in $\mathcal{K}$ is
double trace amplitude of the trace form $\Tr(\cdots)\Tr(t^at^b)$,
and the Feynman diagrams contributing to this amplitude are those
whose $\Phi_{n+1}$ and $\Phi_{n+2}$ are attached to the same double
trace vertex, while the color indices of $\Phi_{n+1},\Phi_{n+2}$ are
separated from others.

Now let us examine the large $z$ behavior of the amplitude under
momentum shifting
$\Spab{\Phi_{n+1}^{\alpha_{n+1}}|\Phi_{n+2}^{\alpha_{n+2}}}$. Since
the color indices of two shifted legs are contracted, it is
equivalent to consider the large $z$ behavior of
$\mathcal{K}(1,2,\ldots, n)$ under such shifting. Following
\cite{Jin:2015pua}, we find that at the large $N$ limit, the leading
interaction part $V$ is given by
\bea V^{\alpha \beta }&=&V^{\alpha \beta}_{\SYM}
+N\bar{\kappa}(\delta^{\alpha}_{\alpha'_{1}}\delta^{\beta}_{\alpha'_{2}}
+\delta^{\alpha}_{\alpha'_{2}}\delta^{\beta}_{\alpha'_{1}})
\bar{\mathcal{O}}+N{\kappa}(T^{\alpha'_{1}\alpha}T^{\alpha'_{2}\beta}
+T^{\alpha'_{2}\alpha}T^{\alpha'_{1}\beta}){\mathcal{O}}~,~~~\label{vnewl}\eea
where $T^{\alpha\beta}$ is defined through
$\Phi^{\alpha}=T^{\alpha\beta}\Phi^{\dagger}_{\beta}$, and
$\alpha'_1=\alpha_{n+1}, \alpha'_2=\alpha_{n+2}$, indicating that
the shifted fields $\Phi_{n+1},\Phi_{n+2}$ are the two fields of
$\Tr(\Phi^{\alpha'_1}\Phi^{\alpha'_2})$ in (\ref{lagrangianO}) with
specific field type. In general, the OPE of shifted fields has the
form \cite{Jin:2015pua}
\bea
\mathcal{Z}(z)=\epsilon^{n+1}_{\alpha}\epsilon^{n+2}_{\beta}\Bigl[V^{\alpha\beta}-V^{\alpha\beta_1}(D_0^{-1})_{\beta_1\beta_2}V^{\beta_2\beta}+\cdots
\Bigr]~,~~~\label{exp-exp-1} \eea
where $\epsilon_\alpha^{n+1},\epsilon_\beta^{n+2}$ are external wave
functions of $\Phi_{n+1},\Phi_{n+2}$. The terms with
$(D_0^{-1})^{k}$ correspond to Feynman diagrams with $k$ hard
propagators. The $\mathcal{Z}(z)$ for $L_{\mathcal{O}}$ contains two
parts, one from the single trace and the other from double trace.
The single trace amplitudes in $\mathcal{K}$ originate from Feynman
diagrams with vertices of $\mathcal{N}=4$ Lagrangian, thus their
$\mathcal{Z}(z)$ can be directly obtained by replacing $V^{\alpha
\beta}$ with $V^{\alpha \beta}_{\SYM}$. The double trace amplitudes
in $\mathcal{K}$ originate from Feynman diagrams with double trace
vertices. Because the two shifted fields $\Phi_{n+1},\Phi_{n+2}$
should be attached to the same double trace vertex, in this case the
hard propagator will not appear in the corresponding Feynman
diagrams. Thus for this part, we only need to keep the first term in
(\ref{exp-exp-1})(more explicitly, the terms with single
$\mathcal{O}$ or $\bar{\mathcal{O}}$ in (\ref{vnewl})). Combined
together, we have
\bea
\mathcal{Z}(z)&=&\mathcal{Z}_{\SYM}(z)+\epsilon^{n+1}_{\alpha}\epsilon^{n+2}_{\beta}N\bar{\kappa}(\delta^{\alpha}_{\alpha'_{1}}\delta^{\beta}_{\alpha'_{2}}
+\delta^{\alpha}_{\alpha'_{2}}\delta^{\beta}_{\alpha'_{1}})\bar{\mathcal{O}}\nonumber\\
&&~~~~~~~~~~~~~~~+\epsilon^{n+1}_{\alpha}\epsilon^{n+2}_{\beta}N{\kappa}(T^{\alpha'_{1}\alpha}T^{\alpha'_{2}\beta}
+T^{\alpha'_{2}\alpha}T^{\alpha'_{1}\beta}){\mathcal{O}}~.~~~\label{zzzz2}\eea
The summation of $\alpha,\beta$ runs over all types of fields. For a
given momentum shifting $\alpha'_1=\alpha_{n+1}$,
$\alpha'_2=\alpha_{n+2}$, we can choose the wave function such that
$\epsilon^{n+1}_{\alpha_{n+1}}\epsilon^{n+2}_{\alpha_{n+2}}\neq 0$
but all other types of contractions vanish. In this case, the second
line of (\ref{zzzz2}) contains a factor
$(T^{\alpha_{n+1}\alpha_{n+1}}T^{\alpha_{n+2}\alpha_{n+2}}
+T^{\alpha_{n+1}\alpha_{n+2}}T^{\alpha_{n+2}\alpha_{n+1}})$. From
the definition of $T^{\alpha\beta}$ in (\ref{Phi2DaggerPhi}), it is
clear that this factor is zero when the two shifted fields are not
complex conjugate to each other. So we have,
\bea \mathcal{Z}(z)=\mathcal{Z}_{\SYM}(z)
+N\bar{\kappa}\epsilon^{n+1}_{\alpha_{n+1}}\epsilon^{n+2}_{\alpha_{n+2}}\bar{\mathcal{O}}~.~~~\label{zzzz3}\eea
However, if the two shifted fields are complex conjugate to each
other, then in the definition of Lagrangian (\ref{lagrangianO}),
$\bar{\mathcal{O}}$ is in fact identical to $\mathcal{O}$. This
means that there is only one term in $\Delta L$ but not two, and
consequently there is only the first line in (\ref{zzzz2}). After
the choice of wave functions, we again get (\ref{zzzz3}).

From eqn.(\ref{zzzz3}), we know that the large $z$ behavior of
$L_{\mathcal{O}}$ under $\spab{\Phi|\Phi}$-shifting depends on the
large $z$ behavior of $\mathcal{N}=4$ SYM theory as well as the
double trace term $\Delta L$. In fact(please refer to Appendix
\ref{largeZN4} for detailed discussion), for all the shifts we use
in this paper\footnote{Including $\langle \phi^I|\phi^J]$, $\langle
\psi^{A\alpha}|\phi^J]$, $\langle
\psi^{A\alpha}|\bar{\psi}_{\dot{\alpha}}]$, $\langle
\psi^{A\alpha}|\psi^{B\beta}]$, $\langle
\psi^{A\alpha}|F^{\beta\gamma}]$, $\langle
\bar{\psi}_{A\dot{\alpha}}|F^{\beta\gamma}]$ and $\langle
\bar{F}^{\dot{\alpha}\dot{\beta}}|F^{\gamma\rho}]$.},
$\mathcal{Z}_{\SYM}(z)$ has lower power in $z$ than the second term
in \eqref{zzzz3} at large $z$. This means that the boundary
operator(or the operator defined by the leading $z$ order) is always
determined by the second term in (\ref{zzzz3}),
\bea
\mathcal{Z}(z)\sim&N\bar{\kappa}\epsilon^{n+1}_{\alpha_{n+1}}\epsilon^{n+2}_{\alpha_{n+2}}\bar{\mathcal{O}}~.~~~
\label{zzzz5}\eea
So it produces the desired operator $\bar{\mathcal{O}}$, up to
certain possible pre-factor from the external wave functions.

%%%%%%%%%%%%%%%
\section{Sudakov form factor and more}
\label{secSudakov}
%%%%%%%%%%%%%%

In this section, we will take the bilinear half-BPS scalar operator
$\mathcal{O}_2\equiv \mathcal{O}_{\I}^{[0]}=\Tr(\phi^I\phi^J)$ as an
example to illustrate the idea of computing form factor from
boundary contributions. The form factor is defined as
\bea \mathcal{F}_{\mathcal{O}_2,n}({s};q)=\int d^4x
e^{-iqx}\Spaa{s|\Tr(\phi^{I}\phi^{J})(x)|0}=\delta^{(4)}(q-\sum_{i=1}^{n}p_i)\Spaa{s|\Tr(\phi^{I}\phi^{J})(0)|0}~.~~~\label{FormO2}\eea
Here $|s\rangle$ is a $n$-particle on-shell states, and each state
in $|s\rangle$ is on-shell, with a momentum $p_i^2=0$, while the
operator, carrying momentum $q=\sum_{i=1}^np_i$, is off-shell. The
simplest example is given by taking
$|s\rangle=|\phi^{I}(p_1)\phi^{J}(p_2)\rangle$, i.e., the Sudakov
form factor, and it is simply\footnote{With coupling constant and
delta function of momentum conservation stripped off here and from
now on for simplicity.}
\bea
\Spaa{\phi^{I}(p_1)\phi^{J}(p_2)|\Tr(\phi^I\phi^J)(0)|0}=1~.~~~\nonumber\eea
A more complicated one is given by taking the on-shell states as two
scalars and $(n-2)$ gluons. Depending on the helicities of gluons,
it defines the MHV form factor, NMHV form factor and so on.

\begin{figure}
  \centering
  \includegraphics[width=6in]{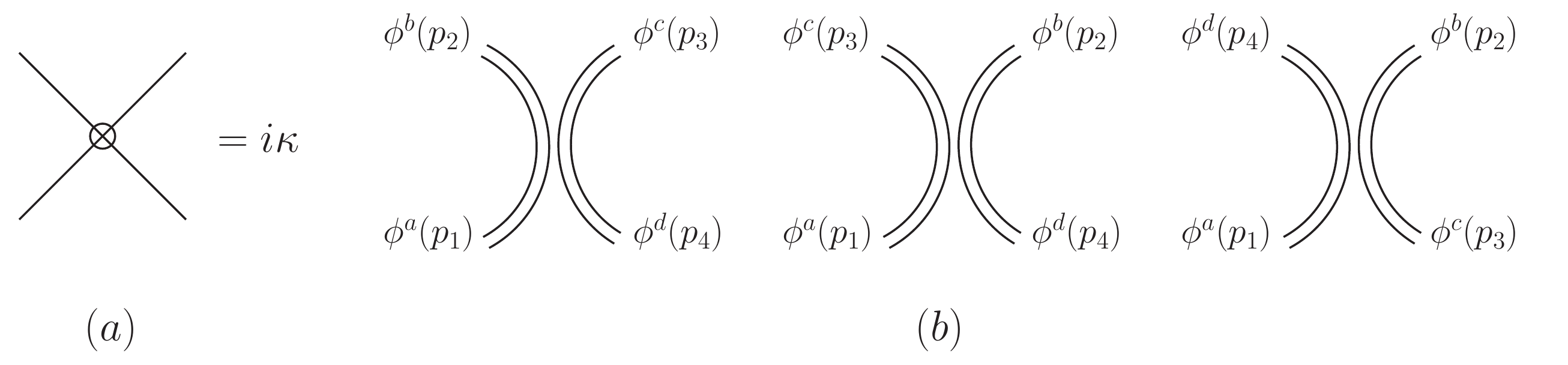}\\
  \caption{(a)The four-scalar vertex of ${\kappa\over 4N}\Tr(\phi^{I}\phi^{J})\Tr(\phi^{K}\phi^{L})$ term, (b)The
  double-line notation of four-scalar vertex, showing the possible trace structures.}\label{FourScalar}
\end{figure}

In order to compute the form factor (\ref{FormO2}) as boundary
contribution of certain amplitude under BCFW shifting, we need to
relate the operator $\mathcal{O}_2$ with certain boundary operator.
This can be done by constructing a new Lagrangian
$L_{\mathcal{O}_2}$ by adding an extra double trace term $\Delta L$
in the $\mathcal{N}=4$ Lagrangian as
\bea L_{\mathcal{O}_2}=L_{\SYM}-{\kappa\over
4N}\Tr(\phi^{I}\phi^{J})\Tr(\phi^{K}\phi^{L})~,~~~\eea
where $\kappa$ is the coupling constant. Since we are dealing with
real scalars, there is no need to add the corresponding complex
conjugate term. This new term provides a four-scalar vertex, and it
equals to $i\kappa$, as shown in Figure (\ref{FourScalar}).
If we split two scalars into ordinary part and hard part
$\phi^{Ia}\to \phi^{Ia}+\phi^{\Lambda Ia}$ and $\phi^{Jb}\to
\phi^{Jb}+\phi^{\Lambda Jb}$(the hard part $\phi^{\Lambda}$
corresponds to the large $z$ part), then the quadratic term
$\phi^{\Lambda Ia}\phi^{\Lambda Jb}$ of $L_{\SYM}$ part can be read
out from the result in Appendix B of \cite{Jin:2015pua} by setting
$A=(A_{\mu},\phi^I)$, which is given by
\bea 2g^2N\delta^{IJ}\Tr(A\cdot
A+\phi\cdot\phi)~.~~~\label{variationSYM}\eea
The quadratic term $\phi^{\Lambda Ia}\phi^{\Lambda Jb}$ of $\Delta
L$ part is simply(at the leading $N$ order)
\bea {N\over 2}\kappa\Tr(\phi^K\phi^L)~.~~~\eea
Thus the boundary operator under two-scalar shifting is
\bea
\mathcal{O}^{\langle\phi^{Ia}|\phi^{Jb}]}=2g^2N\delta^{IJ}\Tr(A\cdot
A+\phi\cdot\phi)+{N\over
2}\kappa\Tr(\phi^K\phi^L)~.~~~\label{O2boundaryO}\eea
Notice that the traceless part (while $I\neq J$) of boundary
operator (\ref{O2boundaryO}) is proportional to the operator
$\mathcal{O}_2$. This means that if the two shifted scalars are not
the same type of scalar, i.e., $I\neq J$, the corresponding boundary
contribution $B^{\langle\phi^{Ia}|\phi^{Jb}]}$ of amplitude defined
by Lagrangian $L_{\mathcal{O}_2}$ is identical to the form factor of
$\mathcal{O}_2=\Tr(\phi^K\phi^L)$, up to some over-all factor which
can be fixed by hand.

More explicitly, let us consider the color-ordered form factor
$\Spaa{1,2,\ldots, n|\mathcal{O}_2|0}$, where $i$ denotes an
arbitrary field. It is dressed with a single trace structure
$\Tr(t^1t^2\cdots t^n)\mathcal{O}_2$. In the amplitude side,
$\mathcal{O}_2$ is generated from the double trace term $\Delta L$,
and the corresponding trace structure of color-ordered amplitude is
$\Tr(t^{1}t^{2}\cdots t^{n})\Tr(t^{n+1}t^{n+2})$. We denote the
amplitude of double trace structure as $A_{n;2}(1,2,\ldots, n;
\phi_{n+1},\phi_{n+2})$. It only gets contributions from the Feynman
diagrams where $\phi_{n+1},\phi_{n+2}$ are attached to the sole
four-scalar vertex of $\Delta L$. Then the form factor
$\Spaa{1,2,\ldots, n|\mathcal{O}_2|0}$ is just the boundary
contribution of $A_{n;2}(1,2,\ldots, n; \phi_{n+1},\phi_{n+2})$
under BCFW shifting of two scalars $\phi_{n+1},\phi_{n+2}$!

As a simple illustration, let us consider four-point scalar
amplitude $A_{2;2}(\phi_1^K,\phi^L_2;\phi^I_3,\phi^J_4)$. In this
case, the only possible contributing diagram is a four-scalar vertex
defined by $ \Delta L$, and we can directly work out as
$A_{2;2}(\phi_1,\phi_2;\phi_3,\phi_4)=i\kappa$. After appropriate
normalization, it can be set as 1. Since it has no dependence on any
external momenta, after momentum shifting
\bea | 3\rangle\to |3\rangle-z|4\rangle~~~,~~~|4]\to
|4]+z|3]~,~~~\eea
the amplitude still remains the same, while the boundary operator is
$\Tr(\phi^K\phi^L)$. There is no pole's term in $z$, while the
zero-th order term in $z$ is $B^{\langle
\phi^I_3|\phi^J_4]}(\phi^K_1,\phi^L_2;\phi^I_{\widehat{3}},\phi^J_{\widehat{4}})=1$.
Thus we confirm the tree-level Sudakov form factor
\bea \Spaa{\phi_1^K,\phi_2^L|\Tr(\phi^K\phi^L)|0}=B^{\langle
3|4]}(\phi_1^K,\phi_2^L;\phi^I_{\widehat{3}},\phi^J_{\widehat{4}})=1~.~~~\eea

Now we have three different ways of studying form factor. The first,
as stated in \cite{Brandhuber:2010ad}, form factor obeys a similar
BCFW recursion relation as amplitude. This enables us to compute a
form factor recursively from lower-point ones. The second, we can
compute the corresponding amplitude. Once it is obtained, we can
take the BCFW shifting $\langle \phi_{n+1}|\phi_{n+2}]$ and extract
the boundary contribution $B^{\spab{\phi_{n+1}|\phi_{n+2}}}$, which
equals to the corresponding form factor after identification. The
third, as stated in \cite{Jin:2014qya}, the boundary contribution
also obeys a similar BCFW recursion relation as amplitude. We can
compute boundary contribution recursively from lower-point boundary
contributions, and once it is obtained, we can work out the form
factor after identification.

In the following subsection, we will take MHV form factor of
operator $\mathcal{O}_2$ as an example, to illustrate these three
ways of understanding.

%%%%%%%%%%%%%%%%%%%%%%%
\subsection{MHV case}
%%%%%%%%%%%%%%%%%%%%%%%%%

%\def\MHV{{\tiny \mbox{MHV}}}
The $n$-point color-ordered MHV form factor of operator
$\mathcal{O}_2$ is given by
\bea
\mathcal{F}^{\MHV}_{\mathcal{O}_2,n}(\{g^+\},\phi_i,\phi_j;q)=-{\Spaa{i~j}^2\over
\Spaa{1~2}\Spaa{2~3}\cdots\Spaa{n~1}}~,~~~\label{MHVformO2}\eea
where
$\mathcal{F}^{\MHV}_{\mathcal{O}_2,n}(\{g^+\},\phi_i,\phi_j;q)$
denotes
$$\mathcal{F}_{\mathcal{O}_2,n}^{\MHV}(g_1^+,\ldots,g_{i-1}^+,\phi_i,g_{i+1}^+,\ldots,
g_{j-1}^+,\phi_j,g_{j+1}^+,\ldots, g_n^+;q)~.$$

%%%%%%%%%%%%%%%%%%%%%%%%%%
\subsubsection*{BCFW recursion relation of form factor}
%%%%%%%%%%%%%%%%%%%%%%%%%%%%

The result (\ref{MHVformO2}) has been proven in paper
\cite{Brandhuber:2010ad}\footnote{Note that we have introduced an
over-all minus sign in the expression (\ref{MHVformO2}), so that the
Sudakov form factor is defined to be
$\mathcal{F}_{\mathcal{O}_2,2}(\phi_1,\phi_2;q)=1$.} by BCFW
recursion relation of form factor. As stated therein, after taking
BCFW shifting of two momenta $p_{i_1},p_{i_2}$, the form factor can
be computed as summation of products of lower-point form factor and
lower-point amplitude, as long as the large $z$ behavior
$\mathcal{F}(z)|_{z\to \infty}\to 0$ is satisfied under such
deformation. The $n$ external legs will be split into two parts,
with $\widehat{p}_{i_1}, \widehat{p}_{i_2}$ in each part separately.
The operator, since it is color-singlet, can be inserted into either
part. So it is possible to build up a $n$-point form factor
recursively from three-point amplitudes and three-point form
factors. Since this method has already been described in
\cite{Brandhuber:2010ad}, we will not repeat it here.

%%%%%%%%%%%%%%%%%%%%%%%%%%
\subsubsection*{BCFW recursion relation of amplitude}
%%%%%%%%%%%%%%%%%%%%%%%%%%%

\begin{figure}
  \centering
  \includegraphics[width=6in]{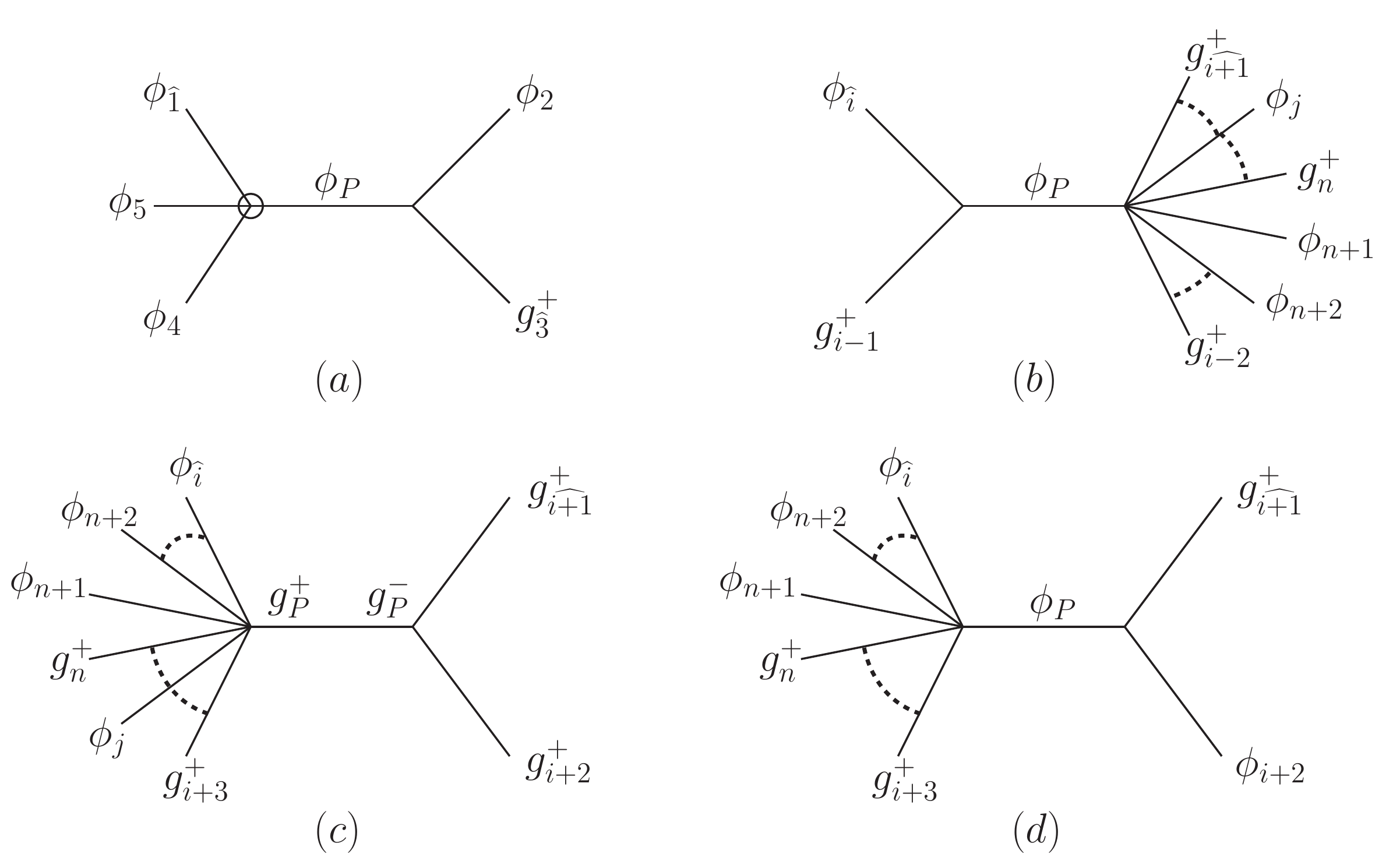}\\
  \caption{(a) is the contributing diagram for $A_{3;2}(\phi_1,\phi_2,g_3^+;\phi_4,\phi_5)$. (b)(c) are the contributing diagrams
  for general $A_{n;2}$ when $j\neq i+2$ while (b)(d) are the contributing diagrams for $A_{n;2}$ when $j=i+2$.}\label{An2proof}
\end{figure}

Instead of computing form factor directly, we can first compute the
corresponding $(n+2)$-point amplitude
\bea A_{n;2}(g_1^+,\ldots,
g_{i-1}^+,\phi_i,g_{i+1}^+,\ldots,g_{j-1}^+,
\phi_j,g_{i+1}^+,\ldots,
g_n^+;\phi_{n+1},\phi_{n+2})~.~~~\label{An2}\eea
This amplitude can be computed via BCFW recursion relation. If we
choose one shifted momentum to be gluon, $A_{n;2}(z)$ will be
vanishing when $z\to \infty$, i.e., there is no boundary
contribution. So we can take $\spab{g^+|\phi}$-shifting in the
computation. The four-point amplitude is trivially
$A_{2;2}(\phi_1,\phi_2;\phi_3,\phi_4)=1$. To compute the five-point
amplitude $A_{3;2}(\phi_1,\phi_2,g_3^+;\phi_4,\phi_5)$, we can take
$\Spab{g^+_3|\phi_1}$-shifting. There is only one contributing term
as shown in Figure (\ref{An2proof}.a), which is given by

\bea
A_{3;2}(\phi_1,\phi_2,g_3^+;\phi_4,\phi_5)&=&A_{2;2}({\phi}_{\widehat{1}},{\phi}_{\widehat{P}};\phi_4,\phi_5){1\over
P_{23}^2}A_3({\phi}_{-\widehat{P}},\phi_2,g^+_{\widehat{3}})\nonumber\\
&=&-1\times{1\over P_{23}^2}\times {\Spbb{2~3}[3~\widehat{P}]\over
[\widehat{P}~2]}=-{\spaa{1~2}^2\over
\spaa{1~2}\spaa{2~3}\spaa{3~1}}~,~~~\eea
where $\widehat{P}=p_2+p_3-z|1\rangle|3]$. Similarly, for general
amplitude $A_{n;2}$, we can take
$\spab{g^+_{i+1}|\phi_{i}}$-shifting\footnote{Because of cyclic
invariance, we can always do this.}. If $j\neq (i+2)$, we need to
consider two contributing terms as shown in Figure
(\ref{An2proof}.b) and (\ref{An2proof}.c), while if $j=(i+2)$, we
need to consider two contributing terms as shown in Figure
(\ref{An2proof}.b) and (\ref{An2proof}.d). In either case,
contribution of diagram (\ref{An2proof}.b) vanishes under
$\Spab{g^+_{i+1}|\phi_i}$-shifting. So we only need to compute
contribution of diagram (\ref{An2proof}.c) or (\ref{An2proof}.d).
Taking $j\neq (i+2)$ as example, we have
\bea &&A_{n;2}(g_1^+,\ldots, \phi_i,\ldots, \phi_j,\ldots,
g_n^+;\phi_{n+1},\phi_{n+2})\\
&=&A_{n-1;2}(g^+_{i+3},\ldots, \phi_j,\ldots, g^+_n,g_1^+,\ldots,
\phi_{\widehat{i}}, g^+_{\widehat{P}};\phi_{n+1},\phi_{n+2}){1\over
P^2_{i+1,i+2}}A_3(g^-_{-\widehat{P}},g^+_{\widehat{i+1}},g^+_{i+2})~.~~~~\nonumber
\eea
Assuming that
\bea
A_{n;2}(\{g^+\},\phi_i,\phi_j;\phi_{n+1},\phi_{n+2})=-{\spaa{i~j}^2\over
\spaa{1~2}\spaa{2~3}\cdots\spaa{n~1}}~~~~\label{an2result}\eea
is true for $A_{n-1;2}$, then
\bea &&A_{n;2}(g_1^+,\ldots, \phi_i,\ldots, \phi_j,\ldots,
g_n^+;\phi_{n+1},\phi_{n+2})\\
&=&-{\spaa{i~j}^2\over \spaa{1~2}\cdots
\spaa{i-1,i}\spaa{i~\widehat{P}}\spaa{\widehat{P},i+3}\spaa{i+3,i+4}\cdots
\spaa{n~1}}{1\over P^2_{i+1,i+2}}{\spbb{i+1,i+2}^3\over
\spbb{\widehat{P},i+1}\spbb{i+2,\widehat{P}}}\nonumber\\
&=&-{\spaa{i~j}^2\over
\spaa{1~2}\spaa{2~3}\cdots\spaa{n~1}}~,~~~\nonumber \eea
where
\bea \widehat{P}=p_{i+1}+p_{i+2}-z_{i+1,i+2}|i\rangle
|i+1]~~,~~z_{i+1,i+2}={\spaa{i+1,i+2}\over \spaa{i,i+2}}~.~~~\eea
Similar computation shows that for $j\neq i+2$ case,
(\ref{an2result}) is also true for all $n$. Thus we have proven the
result (\ref{an2result}) by BCFW recursion relation of amplitude.

As discussed, $\spab{\phi_{n+1}|\phi_{n+2}}$-shifting generates the
boundary operator $\mathcal{O}_2$, and the corresponding boundary
contribution is identical to the form factor of operator
$\mathcal{O}_2$. Here, $A_{n;2}$ does not depend on momenta
$p_{n+1},p_{n+2}$, thus
\bea
B^{\spab{\phi_{n+1}|\phi_{n+2}}}(\{g^+\},\phi_i,\phi_j;\phi_{\widehat{n+1}},\phi_{\widehat{n+2}})=-{\spaa{i~j}^2\over
\spaa{1~2}\spaa{2~3}\cdots\spaa{n~1}}~,~~~\eea
and correspondingly
\bea
\mathcal{F}^{\MHV}_{\mathcal{O}_2,n}(\{g^+\},\phi_i,\phi_j;q)=B^{\spab{\phi_{n+1}|\phi_{n+2}}}=-{\spaa{i~j}^2\over
\spaa{1~2}\spaa{2~3}\cdots\spaa{n~1}}~,~~~\eea
which agrees with the result given by BCFW recursion relation of
form factor.

%%%%%%%%%%%%%%%%%%%%%%%%%
\subsubsection*{Recursion relation of boundary contribution}
%%%%%%%%%%%%%%%%%%%%%%%%%%

We can also compute the boundary contribution directly by BCFW
recursion relation without knowing the explicit expression of
amplitude, as shown in paper \cite{Jin:2014qya}.
\begin{figure}
  \centering
  \includegraphics[width=5.5in]{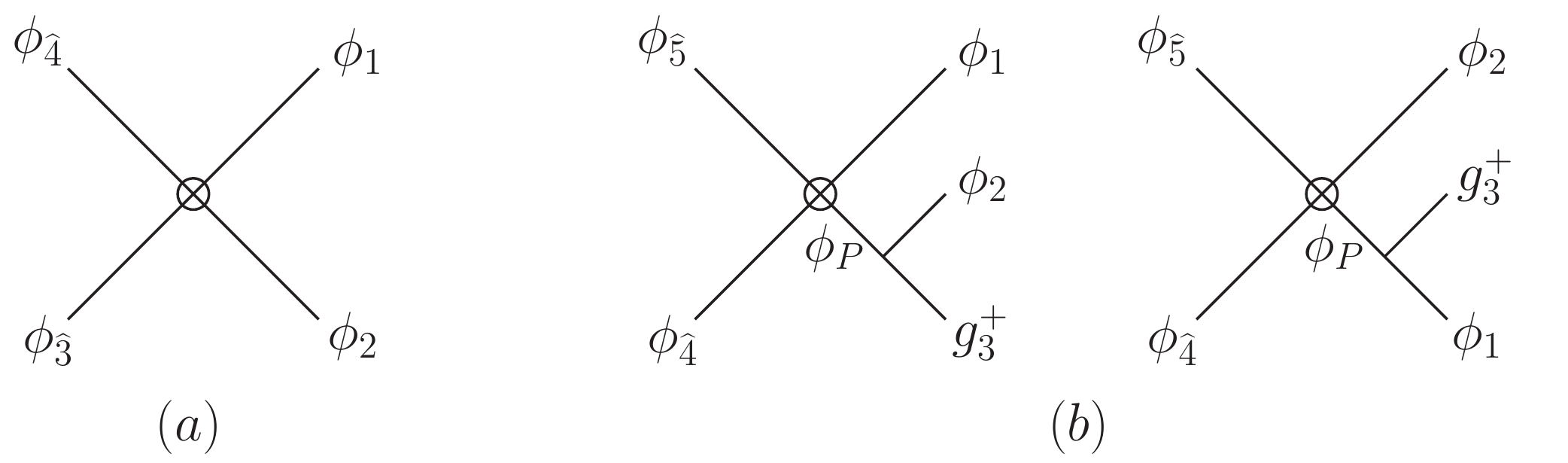}\\
  \caption{(a)Feynman diagram for boundary contribution $B_{2;2}^{\spab{\phi_3|\phi_4}}(\phi_1,\phi_2;\phi_{\widehat{3}},\phi_{\widehat{4}})$, (b)Feynman diagrams for
  boundary contribution $B_{3;2}^{\spab{\phi_4|\phi_5}}(\phi_1,\phi_2,g_3^+;\phi_{\widehat{4}},\phi_{\widehat{5}})$.}\label{Bn2proof}
\end{figure}
The boundary contribution of four and five-point amplitudes can be
computed directly by Feynman diagrams. For four-point case, there is
only one diagram, i.e., four-scalar vertex, as shown in Figure
(\ref{Bn2proof}.a), and
$B^{\spab{\phi_3|\phi_4}}_{2;2}(\phi_1,\phi_2;\phi_{\widehat{3}},\phi_{\widehat{4}})=1$.
For five-point case, under $\spab{\phi_4|\phi_5}$-shifting, only
those Feynman diagrams whose $\widehat{p}_{4},\widehat{p}_{5}$ are
attached to the same four-scalar vertex contribute to the boundary
contribution. There are in total two diagrams as shown in Figure
(\ref{Bn2proof}.b), which gives
\bea
B_{3;2}^{\spab{\phi_4|\phi_5}}(\phi_1,\phi_2,g^+_3;\phi_{\widehat{4}},\phi_{\widehat{5}})&=&-{(p_2-P_{23})^{\mu}\epsilon_{\mu}^{+}(p_3)\over
P_{23}^2}+ {(p_1-P_{13})^{\mu}\epsilon_{\mu}^{+}(p_3)\over
P_{13}^2}\nonumber\\
&=&-{\spaa{1~2}^2\over \spaa{1~2}\spaa{2~3}\spaa{3~1}}~,~~~\eea
where the polarization vector $\epsilon_{\mu}^{\pm}(p)$ is defined
to be
\bea \epsilon_{\mu}^+(p)={\spab{r|\gamma_\mu|p}\over
\sqrt{2}\spaa{r~p}}~~,~~\epsilon_{\mu}^-(p)={\spab{p|\gamma_\mu|r}\over
\sqrt{2}\spbb{p~r}}~,~~~\eea
with $r$ an arbitrary reference spinor. From these lower-point
results, it is not hard to guess that
\bea
B_{n;2}^{\spab{\phi_{n+1}|\phi_{n+2}}}(\{g^+\},\phi_i,\phi_j;\phi_{\widehat{n+1}},\phi_{\widehat{n+2}})=-{\spaa{i~j}^2\over
\spaa{1~2}\spaa{2~3}\cdots\spaa{n~1}}~.~~~\label{bn2result}\eea
This result can be proven recursively by taking another shifting
$\spab{i_1|\phi_{n+2}}$ on $B_{n;2}^{\spab{\phi_{n+1}|\phi_{n+2}}}$,
where $p_{i_1}$ is the momentum other than $p_{n+1},p_{n+2}$. If
under this second shifting, there is no additional boundary
contribution, then $B_{n;2}^{\spab{\phi_{n+1}|\phi_{n+2}}}$ can be
fully determined by the pole terms under
$\spab{i_{1}|\phi_{n+2}}$-shifting. Otherwise we should take a third
momentum shifting and so on, until we have detected the complete
boundary contribution.

Fortunately, if $p_{i_1}$ is the momentum of gluon, a second
shifting $\spab{g_{i_1}^+|\phi_{n+2}}$ is sufficient to detect all
the contributions \cite{Jin:2014qya}. For a general boundary
contribution $B^{\spab{\phi_{n+1}|\phi_{n+2}}}_{n;2}$, we can take
$\spab{g_1^+|\phi_{n+2}}$-shifting. It splits the boundary
contribution into a sub-amplitude times a lower-point boundary
contribution, and only those terms with three-point amplitudes are
non-vanishing. Depending on the location of $\phi_i,\phi_j$, the
contributing terms are different. Assuming that (\ref{bn2result}) is
true for $B_{n-1;2}$, if $i,j\neq 2,n$, we have
\bea
&&B_{n;2}^{\spab{\phi_{n+1}|\phi_{n+2}}}(\{g^+\},\phi_i,\phi_j;\phi_{\widehat{n+1}},\phi_{\widehat{n+2}})\\
&=&A_3(g^+_n,g^+_{\widehat{\widehat{1}}},g^-_{\widehat{\widehat{P}}}){1\over
P_{1n}^2}B^{\spab{\phi_{n+1}|\phi_{n+2}}}_{n-1;2}(g^+_{-\widehat{\widehat{P}}},g^+_2,\ldots,
\phi_i,\ldots, \phi_j,\ldots,
g^+_{n-1};\phi_{\widehat{n+1}},\phi_{\widehat{\widehat{n+2}}})\nonumber\\
&&+A_3(g^+_{\widehat{\widehat{1}}},g^+_2,g^-_{\widehat{\widehat{P}}}){1\over
P_{12}^2}B^{\spab{\phi_{n+1}|\phi_{n+2}}}_{n-1;2}(g^+_{-\widehat{\widehat{P}}},g^+_3,\ldots,
\phi_i,\ldots, \phi_j,\ldots,
g^+_{n};\phi_{\widehat{n+1}},\phi_{\widehat{\widehat{n+2}}})~,~~~\nonumber\eea
while if $i=2,j\neq n$, we have
\bea
&&B_{n;2}^{\spab{\phi_{n+1}|\phi_{n+2}}}(\{g^+\},\phi_2,\phi_j;\phi_{\widehat{n+1}},\phi_{\widehat{n+2}})\\
&=&A_3(g^+_n,g^+_{\widehat{\widehat{1}}},g^-_{\widehat{\widehat{P}}}){1\over
P_{1n}^2}B^{\spab{\phi_{n+1}|\phi_{n+2}}}_{n-1;2}(g^+_{-\widehat{\widehat{P}}},\phi_2,g^+_3,\ldots,
 \phi_j,\ldots,
g^+_{n-1};\phi_{\widehat{n+1}},\phi_{\widehat{\widehat{n+2}}})\nonumber\\
&&~~~~~~+A_3(g^+_{\widehat{\widehat{1}}},\phi_2,\phi_{\widehat{\widehat{P}}}){1\over
P_{12}^2}B^{\spab{\phi_{n+1}|\phi_{n+2}}}_{n-1;2}(\phi_{-\widehat{\widehat{P}}},g^+_3,\ldots,
\phi_j,\ldots,
g^+_{n};\phi_{\widehat{n+1}},\phi_{\widehat{\widehat{n+2}}})~,~~~\nonumber\eea
and if $i=2,j=n$, we have
\bea
&&B_{n;2}^{\spab{\phi_{n+1}|\phi_{n+2}}}(\{g^+\},\phi_2,\phi_n;\phi_{\widehat{n+1},\widehat{n+2}})\\
&=&A_3(\phi_n,g^+_{\widehat{\widehat{1}}},\phi_{\widehat{\widehat{P}}}){1\over
P_{1n}^2}B^{\spab{\phi_{n+1}|\phi_{n+2}}}_{n-1;2}(\phi_{-\widehat{\widehat{P}}},\phi_2,g^+_3,\ldots,
g^+_{n-1};\phi_{\widehat{n+1}},\phi_{\widehat{\widehat{n+2}}})\nonumber\\
&&~~~~~~+A_3(g^+_{\widehat{\widehat{1}}},\phi_2,\phi_{\widehat{\widehat{P}}}){1\over
P_{12}^2}B^{\spab{\phi_{n+1}|\phi_{n+2}}}_{n-1;2}(\phi_{-\widehat{\widehat{P}}},g^+_3,\ldots,g^+_{n-1},\ldots,
\phi_{n};\phi_{\widehat{n+1}},\phi_{\widehat{\widehat{n+2}}})~.~~~\nonumber\eea
All of them lead to the result (\ref{bn2result}), which ends the
proof. Again, with the result of boundary contribution, we can work
out the corresponding form factor directly.

We have shown that the BCFW recursion relation of form factor,
amplitude and boundary contribution lead to the same conclusion.
This is not limited to MHV case, since the connection between form
factor and boundary contribution of amplitude is universal and does
not depend on the external states. In fact, for any form factor with
$n$-particle on-shell states $|s\rangle$, we can instead compute the
corresponding amplitude $A_{n;2}(s;\phi_{n+1},\phi_{n+2})$ defined
by Lagrangian $L_{\mathcal{O}_2}$, and extract the boundary
contribution under $\spab{\phi_{n+1}|\phi_{n+2}}$-shifting. There is
no difference between this boundary contribution and form factor of
$\mathcal{O}_2$. For example, in \cite{Brandhuber:2011tv}, the
authors showed that the split-helicity form factor shares a similar
"zigzag diagram" construction as the split-helicity amplitude given
in \cite{Britto:2005dg}. It is now easy to understand this, since
the form factor is equivalent to the boundary contribution of the
amplitude, and it naturally inherits the "zigzag" construction with
minor modification.

The tree amplitude $A_{n;2}(1,\ldots, n;n+1,n+2)$ associated with
the double trace structure is cyclically invariant inside legs
$\{1,2,\ldots, n\}$ and $\{n+1,n+2\}$, so no surprisingly, the
color-ordered form factor is also cyclically invariant on its $n$
legs. Since the trace structure $\Tr(t^{n+1}t^{n+2})$ is completely
isolated from the other color structure, while the later one is
constructed only from structure constant $f^{abc}$. Thus for
amplitudes $A_{n;2}$, we also have Kleiss-Kuijf(KK)
relation\cite{Kleiss:1988ne} among permutation of legs
$\{1,2,\ldots, n\}$ as
\bea
A_{n;2}(1,\{\alpha\},n,\{\beta\};\phi_{n+1},\phi_{n+2})=(-)^{n_{\beta}}\sum_{\sigma\in
OP\{\alpha\}\cup\{\beta^{T}\}}A_{n;2}(1,\sigma,n;\phi_{n+1},\phi_{n+2})~,~~~\eea
where $n_{\beta}$ is the length of set $\beta$, $\beta^{T}$ is the
reverse of set $\beta$, and $OP$ is the ordered permutation,
containing all the possible permutations between two sets while
keeping each set ordered. This relation can be similarly extended to
form factors. Especially for operator $\mathcal{O}_2$, we can relate
all form factors to those with two adjacent scalars,
\bea
\mathcal{F}_{\mathcal{O}_2,n}(\phi_1,\{\alpha\},\phi_n,\{\beta\};q)=(-)^{n_{\beta}}\sum_{\sigma\in
OP\{\alpha\}\cup\{\beta^{T}\}}\mathcal{F}_{\mathcal{O}_2,n}(\phi_n,\phi_1,\sigma;q)~.~~~\eea
%

%%%%%%%%%%%%%%%%%%%%%%
\subsection{Form factor of operator
$\mathcal{O}_k\equiv\Tr(\phi^{M_1}\phi^{M_2}\cdots \phi^{M_k})$}
%%%%%%%%%%%%%%%%%%

Let us further consider a more general operator $\mathcal{O}_k\equiv
\Tr(\phi^{M_1}\phi^{M_2}\cdots \phi^{M_k})$ and the form factor
$\mathcal{F}_{\mathcal{O}_k,n}(s;q)=\spaa{s|\mathcal{O}_k(0)|0}$. In
order to generate the operator $\mathcal{O}_k$ under certain BCFW
shifting, we need to add an additional Lagrangian term
\bea  \Delta L={\kappa\over (2k)
N}\Tr(\phi^I\phi^J)\Tr(\phi^{M_1}\phi^{M_2}\cdots
\phi^{M_{k}})~~~~\label{FormOk}\eea
to construct a new Lagrangian $L_{\mathcal{O}_{k}}=L_{\SYM}+\Delta
L$. Then the boundary contribution of corresponding amplitude
$A_{n;2}(s;\phi_{n+1},\phi_{n+2})$ under
$\spab{\phi_{n+1}|\phi_{n+2}}$-shifting is identical to the form
factor $\mathcal{F}_{\mathcal{O}_{k},n}(s;q)$.

To see that the boundary operator
$\mathcal{O}^{\spab{\phi^{I_a}|\phi^{J_b}}}$ is indeed the operator
$\mathcal{O}_k$, we can firstly compute the variation of Lagrangian
$L_{\mathcal{O}_k}$ from left with respect to $\phi^{I_a}$, and then
the variation of ${\delta L_{\mathcal{O}_k}\over \delta \phi^{I_a}}$
from right with respect to $\phi^{J_b}$, which we shall denote as
${\overleftarrow{\delta}\over \delta \phi^{J_b}}$ to avoid
ambiguities. The variation of $L_{\SYM}$ part is given in
(\ref{variationSYM}), while for $\Delta L$ part, we have
\bea {\delta \Delta L\over\delta \phi^{I_a} }&=&{\kappa\over
kN}\Tr(\phi^Jt^a)\Tr(\phi^{M_1}\phi^{M_2}\cdots
\phi^{M_{k}})\nonumber\\
&&+{\kappa\over 2N}\Tr(\phi^{N_1}\phi^{N_2})\Tr(t^a
\phi^{M_1}\phi^{M_2}\cdots \phi^{M_{k-1}})~,~~~\eea
and
\bea {\overleftarrow{\delta}\over \delta \phi^{J_b}}\left({\delta
\Delta L\over\delta \phi^{I_a} }\right)&=&{N^2-1\over
2kN}\kappa\Tr(\phi^{M_1}\phi^{M_2}\cdots \phi^{M_k})+{\kappa\over
2N}\phi^{Ma}\Tr(t^a\phi^{M_1}\phi^{M_2}\cdots
\phi^{M_{k-1}})\nonumber\\
&&+\sum_{i}{\kappa\over
2N}\Tr(\phi^{N_1}\phi^{N_2})\Tr(t^a\phi^{M_1}\cdots
\phi^{M_i}t^a\phi^{M_{i+1}}\cdots\phi^{M_{k-2}})~.~~~\nonumber\eea
The first term contains $O(N)$ order result, with a single trace
proportional to $\Tr(\phi^k)$, while the second term is $O({1\over
N})$ order, and the third term is also $O({1\over N})$ order with
even triple trace structure. Thus at the leading $N$ order, the
boundary operator of $L_{\mathcal{O}_k}$ is
\bea
\mathcal{O}^{\langle\phi^{Ia}|\phi^{Jb}]}=2g^2N\delta^{IJ}\Tr(A\cdot
A+\phi^K\phi^K)+{N\over 2k}\kappa\Tr(\phi^{M_1}\phi^{M_2}\cdots
\phi^{M_k})~.~~~\label{OkboundaryO}\eea
Similar to the $\mathcal{O}_2$ case, the traceless part of
(\ref{OkboundaryO}) is proportional to the operator $\mathcal{O}_k$.

The $\Delta L$ term introduces a $(k+2)$-scalar vertex, besides this
it has no difference to $\mathcal{O}_2$ case. We can compute the
amplitude $A_{n;2}(s;\phi_{n+1},\phi_{n+2})$, take
$\spab{\phi_{n+1}|\phi_{n+2}}$-shifting and extract the boundary
contribution. Then transforming it to form factor is almost trivial.
For instance,
$A_{k;2}(\phi_1,\ldots,\phi_k;\phi_{k+1},\phi_{k+2})=1$, thus
$\mathcal{F}_{\mathcal{O}_k,k}(\phi_1,\ldots,\phi_k;q)=1$. It is
also easy to conclude that, since the Feynman diagrams of amplitude
$$A_{n;2}(\phi_1,\cdots,\phi_{k},g_{k+1}^+,\ldots,
g_{n}^+;\phi_{n+1},\phi_{n+2})$$ defined by $L_{\mathcal{O}_k}$ have
one-to-one mapping to the Feynman diagrams of amplitude
$$A_{n-(k-2);2}(\phi_1,\phi_{k},g_{k+1}^+,\ldots,
g_n^+;\phi_{n+1},\phi_{n+2})$$ defined by $L_{\mathcal{O}_2}$ by
just replacing the $(k+2)$-scalar vertex with four-scalar vertex, we
have
\bea
A^{\mathcal{O}_k}_{n;2}(\phi_1,\ldots,\phi_{k},g_{k+1}^+,\ldots,g_n^+;\phi_{n+1},\phi_{n+2})&=&A^{\mathcal{O}_2}_{n-(k-2);2}(\phi_1,\phi_{k},g_{k+1}^+,\ldots,g_n^+;\phi_{n+1},\phi_{n+2})\nonumber\\
&=&-{\spaa{1~k}\over\spaa{k,k+1}\spaa{k+1,k+2}\cdots\spaa{n~1}}~.~~~\eea
Thus we get
\bea
\mathcal{F}_{\mathcal{O}_k;n}(\phi_1,\ldots,\phi_{k},g_{k+1}^+,\ldots,g_n^+;q)&=&-{\spaa{1~k}\over\spaa{k,k+1}\spaa{k+1,k+2}\cdots\spaa{n~1}}~.~~~\eea
%

%%%%%%%%%%%%%%%%%%%%%%%%%%%
\section{Form factor of composite operators}
\label{secComposite}
%%%%%%%%%%%%%%%%%%%%%%%%%%%%%%

Now we move to the computation of form factors for the composite
operators introduced in \S \ref{secOperator}. For convenience we
will use complex scalars $\phi^{AB},\bar{\phi}_{AB}$ instead of real
scalars $\phi^I$ in this section. We will explain the construction
of Lagrangian which generates the corresponding operators, and
compute the MHV form factors through amplitudes of double trace
structure.

%%%%%%%%%%%%%%%%%%%%%%
\subsection{The spin-0 operators}
%%%%%%%%%%%%%%%%%%%%%%%%%

%\def\I{\tiny \mbox{I}}
%\def\II{\tiny \mbox{II}}
%\def\III{\tiny \mbox{III}}
%\def\IX{\tiny \mbox{IX}}
%\def\X{\tiny \mbox{X}}
%\def\XI{\tiny \mbox{XI}}
There are three operators
\bea
&&\mathcal{O}^{[0]}_{\I}=\Tr(\phi^{AB}\phi^{CD})~~~,~~~\mathcal{O}^{[0]}_{\II}=\Tr(\psi^{A\gamma}\psi^B_{\gamma})~~~,~~~\mathcal{O}^{[0]}_{\III}=\Tr(F^{\alpha\beta}F_{\alpha\beta})~,~~~\eea
with their complex conjugate partners
$\bar{\mathcal{O}}_{\I}^{[0]}$, $\bar{\mathcal{O}}_{\II}^{[0]}$ and
$\bar{\mathcal{O}}_{\III}^{[0]}$. For these operators, in order to
construct Lorentz invariant double trace Lagrangian terms $\Delta
L$, we need to product them with another spin-0 trace term. Since
shifting a gluon is always more complicated than shifting a fermion,
and shifting a fermion is more complicated than shifting a scalar,
we would like to choose the spin-0 trace term as trace of two
scalars, as already shown in operator $\mathcal{O}_2$ case.

For operator $\mathcal{O}_{\II}^{[0]}$, we could construct the
Lagrangian as
\bea L_{\mathcal{O}^{[0]}_{\II}}=L_{\SYM}+{\kappa \over
N}\Tr(\phi^{A'B'}\phi^{C'D'})\Tr(\psi^{A\gamma}\psi^B_{\gamma})+{\bar{\kappa}\over
N}
\Tr(\bar{\phi}_{A'B'}\bar{\phi}_{C'D'})\Tr(\bar{\psi}^{\dot{\gamma}}_A\bar{\psi}_{B\dot{\gamma}})~.~~~\eea
The momentum shifting of two scalars $\phi_{n+1},\phi_{n+2}$ will
generate the boundary operator
$\mathcal{O}^{\spab{\phi_{n+1}|\phi_{n+2}}}=\Tr(\bar{\psi}^{\dot{\gamma}}_A\bar{\psi}_{B\dot{\gamma}})$,
while the shifting of two scalars
$\bar{\phi}_{n+1},\bar{\phi}_{n+2}$ will generate the boundary
operator
$\mathcal{O}^{\spab{\bar{\phi}_{n+1}|\bar{\phi}_{n+2}}}=\Tr(\psi^{A\gamma}\psi^B_{\gamma})$.
Thus the form factor
$$\mathcal{F}_{\mathcal{O}^{[0]}_{\II},n}(s;q)=\spaa{s|\mathcal{O}^{[0]}_{\II}|0}$$
is identical to the boundary contribution of amplitude
$A_{n;2}(s;\bar{\phi}_{n+1},\bar{\phi}_{n+2})$ defined by
$L_{\mathcal{O}^{[0]}_{\II}}$ under
$\spab{\bar{\phi}_{n+1}|\bar{\phi}_{n+2}}$-shifting. This amplitude
can be computed by Feynman diagrams or BCFW recursion relation
method.

The $\Delta L$ Lagrangian term introduces
$\phi$-$\phi$-$\psi$-$\psi$ and
$\bar{\phi}$-$\bar{\phi}$-$\bar{\psi}$-$\bar{\psi}$ vertices in the
Feynman diagrams, and it defines the four-point amplitude
$A_{2;2}(\bar{\psi}_1,\bar{\psi}_2;\bar{\phi}_3,\bar{\phi}_4)=\spaa{1~2}$
as well as $A_{2;2}(\psi_1,\psi_2;\phi_3,\phi_4)=\spbb{1~2}$. Thus
it is immediately know that the boundary contribution
$B^{\spab{\bar{\phi}_3|\bar{\phi}_4}}(\bar{\psi}_1,\bar{\psi}_2;\bar{\phi}_{\widehat{3}},\bar{\phi}_{\widehat{4}})=\spaa{1~2}$,
and the form factor
$\mathcal{F}_{\mathcal{O}^{[0]}_{\II},n}(\bar{\psi}_1,\bar{\psi}_2;q)=\spaa{1~2}$.
We can also compute the five-point amplitude
$A_{3;2}(\bar{\psi}_1,\bar{\psi}_2,g_3^+;\bar{\phi}_4,\bar{\phi}_5)$,
and the contributing Feynman diagrams are similar to Figure
(\ref{Bn2proof}.b) but now we have $\bar{\psi}_1,\bar{\psi}_2$
instead of $\bar{\phi}_1,\bar{\phi}_2$. It is given by
\bea
A_{3;2}(\bar{\psi}_1,\bar{\psi}_2,g_3^+;\bar{\phi}_4,\bar{\phi}_5)&=&{\spaa{1|P_{23}|\gamma^{\mu}|2}\over
s_{23}}\epsilon^{+}_{\mu}(p_3)+{\spaa{2|P_{13}|\gamma^{\mu}|1}\over
s_{13}}\epsilon^{+}_{\mu}(p_3)=-{\spaa{1~2}^2\over
\spaa{2~3}\spaa{3~1}}~.~~~\eea
Generalizing this result to $(n+2)$-point double trace amplitude, we
have
\bea
A_{n;2}(\{g^+\},\bar{\psi}_i,\bar{\psi}_j;\bar{\phi}_{n+1},\bar{\phi}_{n+2})=-{\spaa{i~j}^3\over
\spaa{1~2}\spaa{2~3}\spaa{3~4}\cdots
\spaa{n~1}}~.~~~\label{fermionMHVForm}\eea
It is easy to verify above result by BCFW recursion relation of
amplitude, for example, by taking
$\spab{g_1^+|\bar{\psi}_i}$-shifting. Similar to the $\mathcal{O}_2$
case, only those terms with three-point sub-amplitudes can have
non-vanishing contributions, and after substituting the explicit
results for $A_3$ and $A_{n-1;2}$, we arrive at the result
(\ref{fermionMHVForm}). The boundary contribution of amplitude
(\ref{fermionMHVForm}) under
$\spab{\bar{\phi}_{n+1}|\bar{\phi}_{n+2}}$-shifting keeps the same
as $A_{n;2}$ itself, thus consequently we get the form factor
\bea
\boxed{\mathcal{F}_{\mathcal{O}^{[0]}_{\II},n}(\{g^+\},\bar{\psi}_i,\bar{\psi}_j;q)=-{\spaa{i~j}^3\over
\spaa{1~2}\spaa{2~3}\spaa{3~4}\cdots \spaa{n~1}}~.~~~}\eea
It is also interesting to consider another special $n$-point
external states, i.e., two fermions with $(n-2)$ gluons of negative
helicities. For five-point amplitude
$A_{3;2}(\bar{\psi}_1,\bar{\psi}_2,g_3^-;\bar{\phi}_4,\bar{\phi}_5)$,
the contributing Feynman diagrams can be obtained by replacing
$g_3^+$ as $g_3^-$ in amplitude
$A_{3;2}(\bar{\psi}_1,\bar{\psi}_2,g_3^+;\bar{\phi}_4,\bar{\phi}_5)$,
so we have
\bea
A_{3;2}(\bar{\psi}_1,\bar{\psi}_2,g_3^-;\bar{\phi}_4,\bar{\phi}_5)&=&{\spaa{1|P_{23}|\gamma^{\mu}|2}\over
s_{23}}\epsilon^{-}_{\mu}(p_3)+{\spaa{2|P_{13}|\gamma^{\mu}|1}\over
s_{13}}\epsilon^{-}_{\mu}(p_3)\nonumber\\
&=&{(p_4+p_5)^2\spbb{1~2}\over
\spbb{1~2}\spbb{2~3}\spbb{3~1}}~.~~~\eea
More generally, we have
\bea
A_{n;2}(\{g^-\},\bar{\psi}_i,\bar{\psi}_j;\bar{\phi}_{n+1},\bar{\phi}_{n+2})={(p_{n+1}+p_{n+2})^2\spbb{i~j}\over
\spbb{1~2}\spbb{2~3}\cdots
\spbb{n~1}}~.~~~\label{fermionAllMinus}\eea
This result can be proven recursively by BCFW recursion relation.
Assuming eqn. (\ref{fermionAllMinus}) is valid for $A_{n-1;2}$, then
taking $\spab{\bar{\phi}_{n+2}|g_n}$-shifting, we get two
contributing terms\footnote{We assumed that $i,j\neq 1,n-1$,
otherwise the two contributing terms are slightly different. However
they lead to the same conclusion.} for $A_{n;2}$. The first term is
\bea &&A_3(g^-_{\widehat{n}},g^-_1,g^+_{\widehat{P}_{1n}}){1\over
P_{1n}^2}A_{n-1;2}(g^-_{-\widehat{P}_{1n}},g^-_2,\ldots,
\bar{\psi}_i,\ldots,\bar{\psi}_j,\ldots,g^-_{n-1};\bar{\phi}_{n+1},\bar{\phi}_{\widehat{n+2}})\nonumber\\
&=&{\spbb{i~j}(p_{n+1}+p_{n+2})^2\over
\spbb{1~2}\spbb{2~3}\cdots\spbb{n-1,n}\spbb{n~1}}{\spbb{n+2,1}\spbb{n,n-1}\over
\spbb{n-1,1}\spbb{n+2,n}}\nonumber\\
&&~~~~+{\spbb{i~j}\over
\spbb{1~2}\spbb{2~3}\cdots\spbb{n-1,n}\spbb{n~1}}{\spaa{n+1,n}\spbb{n+2,n+1}\over
\spbb{n-1,1}\spbb{n+2,n}}\spbb{n~1}\spbb{n,n-1}~,~~~\eea
while the second term is
%$$|\widehat{n}]=|n]-z|n+2]~~~,~~~|\widehat{n+2}\rangle=|n+2\rangle+z|n\rangle~,$$
%For this one $\widehat{P}_{1n}=p_1+p_n-z|n\rangle|n+2]$,
%$z_{1n}={\spbb{n~1}\over \spbb{n+2,1}}$, and
%$$(p_{n+1}+\widehat{p}_{n+2})^2=(p_{n+1}+p_{n+2})^2+\spaa{n+1,n}\spbb{n+2,n+1}{\spbb{n~1}\over \spbb{n+2,1}}~,$$ so we have
%
\bea
&&A_3(g_{n-1}^-,g^-_{\widehat{n}},g^{+}_{\widehat{P}_{n-1,n}}){1\over
P_{n-1,n}^2}A_{n-1;2}(g^{-}_{-\widehat{P}_{n-1,n}},g^-_1,\ldots,\bar{\psi}_i,\ldots,\bar{\psi}_j,\ldots,g_{n-2}^-;\bar{\phi}_{n+1},\bar{\phi}_{\widehat{n+2}})\nonumber\\
&=&{\spbb{i~j}(p_{n+1}+p_{n+2})^2\over
\spbb{1~2}\spbb{2~3}\cdots\spbb{n-1,n}\spbb{n~1}}{\spbb{n-1,n+2}\spbb{n,1}\over
\spbb{n-1,1}\spbb{n+2,n}}\nonumber\\
&&~~~~+{\spbb{i~j}\over
\spbb{1~2}\spbb{2~3}\cdots\spbb{n-1,n}\spbb{n~1}}{\spaa{n+1,n}\spbb{n+2,n+1}\over
\spbb{n-1,1}\spbb{n+2,n}}\spbb{n~1}\spbb{n-1,n}~.~~~\eea
%
%For this one, $\widehat{P}_{n-1,n}=p_{n-1}+p_n-z|n\rangle|n+2]$,
%$z_{n-1,n}={\spbb{n,n-1}\over \spbb{n+2,n-1}}$, and
%$$(p_{n+1}+\widehat{p}_{n+2})^2=(p_{n+1}+p_{n+2})^2+\spaa{n+1,n}\spbb{n+2,n+1}{\spbb{n,n-1}\over \spbb{n+2,n-1}}~,$$ so we have
%
Summing above two contributions, we get the desired eqn.
(\ref{fermionAllMinus}).

Note that $q=-p_{n+1}-p_{n+2}$ shows up in result
(\ref{fermionAllMinus}), which is the momentum carried by the
operator in form factor. The
$\spab{\bar{\phi}_{n+1}|\bar{\phi}_{n+2}}$-shifting assures that
$\widehat{p}_{n+1}+\widehat{p}_{n+2}=p_{n+1}+p_{n+2}$, thus we get
the form factor
\bea
\boxed{\mathcal{F}_{\mathcal{O}_{\II}^{[0]},n}(\{g^-\},\bar{\psi}_i,\bar{\psi}_j;q)={q^2\spbb{i~j}\over
\spbb{1~2}\spbb{2~3}\cdots\spbb{n~1}}~.~~~}\eea

\begin{figure}
\centering
  \includegraphics[width=6in]{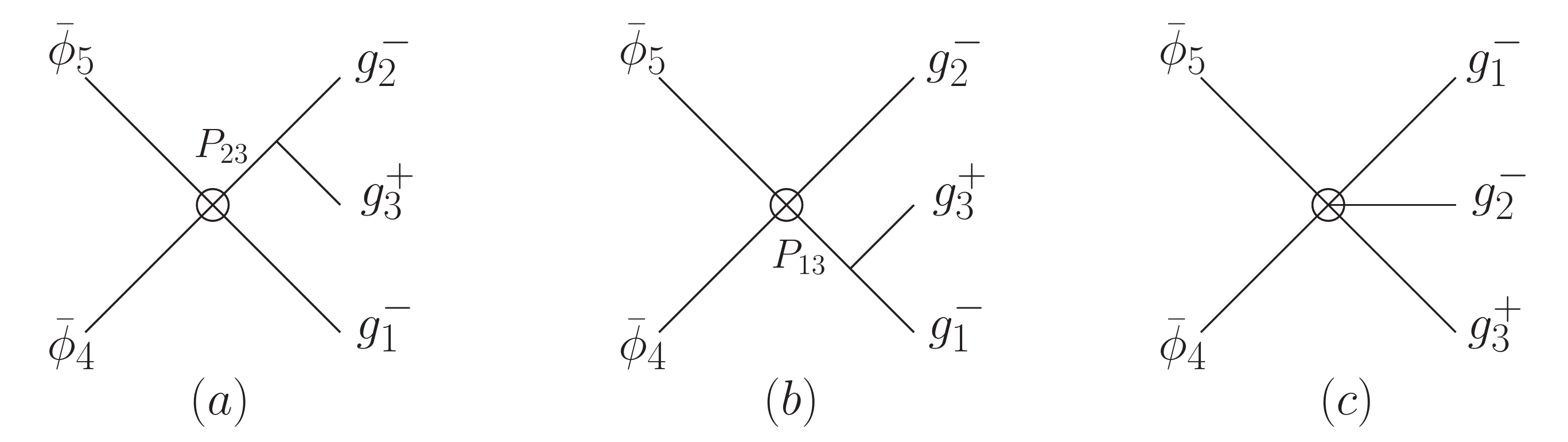}\\
  \caption{Feynman diagrams for $A_{3;2}(g_1^-,g_2^-,g_3^+;\bar{\phi}_4,\bar{\phi}_5)$
  defined by $L_{\mathcal{O}^{[0]}_{\III}}$. All external particles are out-going.}\label{ssgmgmgp}
\end{figure}

For operator $\mathcal{O}_{\III}^{[0]}$, we can also construct the
Lagrangian as
\bea L_{\mathcal{O}^{[0]}_{\III}}=L_{\SYM}+{\kappa \over
N}\Tr(\phi^{A'B'}\phi^{C'D'})\Tr(F^{\alpha\beta}F_{\alpha\beta})+{\bar{\kappa}\over
N}
\Tr(\bar{\phi}_{A'B'}\bar{\phi}_{C'D'})\Tr(\bar{F}^{\dot{\alpha}\dot{\beta}}\bar{F}_{\dot{\alpha}\dot{\beta}})~.~~~\eea
As usual, the $\spab{\bar{\phi}_{n+1}|\bar{\phi}_{n+2}}$-shifting
generates the boundary operator
$\mathcal{O}^{\spab{\bar{\phi}_{n+1}|\bar{\phi}_{n+2}}}=\Tr(F^{\alpha\beta}F_{\alpha\beta})$,
while the $\Delta L$ double trace Lagrangian term introduces four,
five and six-point vertices in the Feynman diagrams. For
computational convenience, let us take the following definition of
self-dual $F^{+}_{\mu\nu}$ and anti-self-dual $F^{-}_{\mu\nu}$ field
strengthes
\bea F_{\mu\nu}^{\pm}={1\over 2}F_{\mu\nu}\pm{1\over
4i}\epsilon_{\mu\nu\rho\sigma}F^{\rho\sigma}~~\mbox{and}~~{1\over
2}\epsilon_{\mu\nu\rho\sigma}F^{\pm\rho\sigma}=\pm
F^{\pm}_{\mu\nu}~,~~~\eea
and rewrite the Lagrangian as
\bea L_{\mathcal{O}^{[0]}_{\III}}=L_{\SYM}+\kappa
\Tr(\phi^{A'B'}\phi^{C'D'})\Tr(F^{+\mu\nu}F^{+}_{\mu\nu})+\bar{\kappa}
\Tr(\bar{\phi}_{A'B'}\bar{\phi}_{C'D'})\Tr(F^{-\mu\nu}F^{-}_{\mu\nu})~.~~~\nonumber\eea
The off-shell Feynman rules for the four-point vertices defined by
the corresponding terms inside $\Tr(\phi\phi)\Tr(F^{+}F^{+})$ or
$\Tr(\bar{\phi}\bar{\phi})\Tr(F^{-}F^{-})$ of $\Delta L$ are given
by
\bea M_{\mu\nu}^{\pm}=(p_{i_1}\cdot
p_{i_2})\eta_{\mu\nu}-p_{i_1\nu}p_{i_2\mu}\pm{1\over
i}\epsilon_{\mu\nu\rho\sigma}p_{i_1}^{\rho}p_{i_1}^{\sigma}~,~~~\eea
where $p_{i_1},p_{i_2}$ are the momenta of two gluons. In fact,
$M^{+}_{\mu\nu}$ can only attach gluons with positive helicities
while $M^{-}_{\mu\nu}$ can only attach gluons with negative
helicities, since
\bea \epsilon_1^{+\mu}M_{\mu\nu}^+={\spbb{1|\gamma_{\nu}|p_2|1}\over
\sqrt{2}}~~,~~\epsilon^{-\mu}_1M_{\mu\nu}^+=0~~\mbox{and}~~\epsilon_1^{+\mu}M_{\mu\nu}^-=0~~,~~\epsilon_1^{-\mu}M_{\mu\nu}^-={\spaa{1|\gamma_\nu|p_2|1}\over
\sqrt{2}}~.~~~\nonumber\eea
And the four-point amplitudes defined by these vertices are given by
\bea
&&A_{2;2}(g_1^-,g_2^-;\bar{\phi}_3,\bar{\phi}_4)=\epsilon_1^{-\mu}M_{\mu\nu}^-\epsilon_2^{-\nu}=\spaa{1~2}^2~~,~~A_{2;2}(g_1^+,g_2^+;\phi_3,\phi_4)=\epsilon_1^{+\mu}M_{\mu\nu}^+\epsilon_2^{+\nu}=\spbb{1~2}^2~.~~~\nonumber\eea
In order to compute the five-point amplitude
$A_{3;2}(g_1^-,g_2^-,g_3^+;\bar{\phi}_4,\bar{\phi}_5)$, we also need
the Feynman rule for five-point vertex defined by the corresponding
terms inside $\Tr(\phi\phi)\Tr(F^{+}F^{+})$ or
$\Tr(\bar{\phi}\bar{\phi})\Tr(F^{-}F^{-})$, which is given by
\bea &&V^{abc}_{\mu\nu\rho}\label{fivepointV}\\
&&={ig\over
2}f^{abc}\Big((p_1-p_2)_{\rho}\eta_{\mu\nu}+(p_2-p_3)_{\mu}\eta_{\nu\rho}+
(p_3-p_1)_\nu\eta_{\rho\mu}+i\kappa(p_1+p_2+p_3)^{\sigma}\epsilon_{\mu\nu\sigma\rho}\Big)~.~~~\nonumber\eea
There are in total three contributing Feynman diagrams, as shown in
Figure (\ref{ssgmgmgp}). We need to sum up all of three results.
The first diagram gives
\bea (a)&=&{\spaa{1|P_{23}|\gamma_{\mu}|1}\over
P_{23}^2}\Big((\epsilon_3^+\cdot\epsilon_2^-)p_2^{\mu}-(P_{23}\cdot
\epsilon_3)\epsilon_2^{-\mu}+(p_3\cdot
\epsilon_2^-)\epsilon_3^{+\mu}\Big)\nonumber\\
&=&{\spaa{r_3~2}\spaa{1~2}^2\over
\spaa{2~3}\spaa{3~r_3}}+{\spaa{1~2}\spaa{r_3~1}\spbb{r_2~3}\over
\spaa{r_3~3}\spbb{2~r_2}}~,~~~\eea
where $r_1,r_2,r_3$ are reference momenta of
$\epsilon_{\mu}^-(p_1),\epsilon_{\mu}^-(p_2),\epsilon_{\mu}^+(p_3)$(abbreviate
as $\epsilon_1^-$, $\epsilon_2^-$, $\epsilon_3^+$) respectively. The
second diagram gives
\bea (b)&=&{\spaa{2|P_{13}|\gamma_{\mu}|2}\over
P_{13}^2}\Big(-(P_{13}\cdot\epsilon_1^-)\epsilon_3^{+\mu}+(p_1\cdot\epsilon_3^+)\epsilon_1^{-\mu}+(\epsilon_1^-\epsilon_3^+)p_3^\mu\Big)\nonumber\\
&=&{\spaa{1~r_3}\spaa{1~2}^2\over\spaa{3~1}\spaa{r_3~3}}+{\spaa{1~2}\spaa{r_3~2}\spbb{r_1~3}\over
\spaa{r_3~3}\spbb{1~r_1}}~.~~~\eea
The third diagram (\ref{ssgmgmgp}.c) is defined by the five-point
vertex (\ref{fivepointV}), while the result of first three terms in
the bracket of (\ref{fivepointV}) is
\bea (c.1)&=&{1\over 2}((p_2-p_1)\cdot
\epsilon_3^+)(\epsilon_1^-\cdot\epsilon_2^-)+((p_1-p_3)\cdot
\epsilon_2^-)(\epsilon_3^+\cdot\epsilon_1^-)+((p_3-p_2)\cdot
\epsilon_1^-)(\epsilon_2^-\cdot\epsilon_3^+)\nonumber\\
&=&{1\over
2}\Big({\spbb{r_1~3}\spaa{1~2}\spaa{2~r_3}\over\spaa{r_3~3}\spbb{1~r_1}}-{\spbb{3~r_2}\spaa{1~2}\spaa{1~r_3}\over\spaa{r_3~3}\spbb{2~r_2}}+{\spbb{r_1~3}\spbb{r_2~3}\spaa{2~1}\over\spbb{1~r_1}\spbb{2~r_2}}\Big)~.~~~\eea
Using
\bea
i\epsilon_{\mu\nu\rho\sigma}p_1^{\mu}p_2^{\nu}p_3^{\rho}p_4^{\sigma}=\spaa{1~2}\spbb{2~3}\spaa{3~4}\spbb{4~1}-\spbb{1~2}\spaa{2~3}\spbb{3~4}\spaa{4~1}~,~~~\nonumber\eea
the last term in the bracket of (\ref{fivepointV}) can be computed
as
\bea
(c.2)&=&{1\over 2}i\epsilon_{\mu\nu\sigma\rho}\epsilon_1^{-\mu}\epsilon_2^{-\nu}(p_1+p_2+p_3)^\sigma\epsilon_3^{+\rho}\nonumber\\
&=&{1\over 2}\Big({\spaa{1~2}\spaa{2~r_3}\spbb{r_1~3}\over
\spbb{1~r_1}\spaa{r_3~3}}-{\spaa{1~2}\spaa{1~r_3}\spbb{3~r_2}\over
\spbb{2~r_2}\spaa{r_3~3}}+{\spaa{1~2}\spbb{r_2~3}\spbb{r_1~3}\over
\spbb{1~r_1}\spbb{2~r_2}}\Big)~.~~~\eea
Summing above contributions, we get
\bea A_{3;2}(g_1^-,g_2^-,g_3^+;\bar{\phi}_4,\bar{\phi}_5)
=-{\spaa{1~2}^4\over \spaa{1~2}\spaa{2~3}\spaa{3~1}}~.~~~\eea
More generally, we have
\bea A_{n;2}(\{g^+\},
g_i^-,g_j^-;\bar{\phi}_{n+1},\bar{\phi}_{n+2})=-{\spaa{i~j}^4\over
\spaa{1~2}\spaa{2~3}\cdots\spaa{n~1}}~,~~~\eea
which can be trivially proven by BCFW recursion relation. This
expression is exactly the same as the pure-gluon $n$-point MHV
amplitude of Yang-Mills theory. By taking
$\spab{\bar{\phi}_{n+1}|\bar{\phi}_{n+2}}$-shifting, we can get the
form factor as
\bea
\boxed{\mathcal{F}_{\mathcal{O}^{[0]}_{\III},n}(\{g^+\},g_i^-,g_j^-;q)=-{\spaa{i~j}^4\over\spaa{1~2}\spaa{2~3}\cdots\spaa{n~1}}~.~~~}\eea

Again, let us consider another configuration of external states,
i.e., $n$ gluons with negative helicities and two scalars.
Computation of
$A_{3;2}(g_1^-,g_2^-,g_3^-;\bar{\phi}_4,\bar{\phi}_5)$ is almost the
same as $A_{3;2}(g_1^-,g_2^-,g_3^+;\bar{\phi}_4,\bar{\phi}_5)$, and
we only need to replace $\epsilon_3^{+}$ by $\epsilon_3^{-}$. Direct
computation shows that, contributions of all three diagrams lead to
\bea
{s_{12}^2+s_{13}^2+s_{23}^2+2s_{12}s_{13}+2s_{12}s_{23}+2s_{13}s_{23}\over\spbb{1~2}\spbb{2~3}\spbb{3~1}}={((p_4+p_5)^2)^2\over
\spbb{1~2}\spbb{2~3}\spbb{3~1}}~.~~~\eea
This result can be generalized to $A_{n;2}$ as
\bea A_{n;2}(g_1^-,g_2^-,\ldots,
g_n^-;\bar{\phi}_{n+1},\bar{\phi}_{n+2})={((p_{n+1}+p_{n+2})^2)^2\over
\spbb{1~2}\spbb{2~3}\cdots\spbb{n~1}}~,~~~\label{gluonAllminus}\eea
and can be proven recursively by BCFW recursion relation. In fact,
assuming eqn. (\ref{gluonAllminus}) is true for $A_{n-1;2}$ and
taking $\spab{g_n^-|g_1^-}$-shifting, there is only one
non-vanishing term in BCFW expansion, which gives
\bea &&A_3(g_{\widehat{1}}^-,g_2^-,g_{\widehat{P}_{12}}^+){1\over
P_{12}^2}A_{n-1;2}(g_{-\widehat{P}_{12}}^-,g_3^-,\ldots,
g_{\widehat{n}}^-;\bar{\phi}_{n+1},\bar{\phi}_{n+2})={((p_{n+1}+p_{n+2})^2)^2\over
\spbb{1~2}\spbb{2~3}\cdots\spbb{n~1}}~.~~~\eea
So the corresponding form factor is
\bea
\boxed{\mathcal{F}_{\mathcal{O}_{\III}^{[0]},n}(g_1^-,g_2^-,\ldots,
g_n^-;q)={(q^2)^2\over \spbb{1~2}\spbb{2~3}\cdots
\spbb{n~1}}~.~~~}\eea
%

%%%%%%%%%%%%%%%%%%%%%%%%%%
\subsection{The spin-${1\over 2}$ operators}
%%%%%%%%%%%%%%%%%%%%%%%%%

For operators
\bea
\mathcal{O}^{[{1/2}]}_{\I}=\Tr(\phi^{AB}\psi^{C\alpha})~~~~~~~~,~~~~~~~~\mathcal{O}^{[{1/2}]}_{\II}=\Tr(\psi^{A}_{\beta}F^{\beta\alpha})~,~~~\eea
and their complex conjugates $\bar{\mathcal{O}}^{[1/2]}_{\I},
\bar{\mathcal{O}}^{[1/2]}_{\II}$, we need to product them with
another spin-${1\over 2}$ trace term, which can be chosen as trace
of product of scalar and fermion.

For operator ${\mathcal{O}}^{[1/2]}_{\I}$, we can construct the
Lagrangian as
\bea L_{\mathcal{O}_{\I}^{[{1/2}]}}=L_{\SYM}+{\kappa\over
N}\Tr(\phi^{A'B'}\psi^{C'\alpha})\Tr(\phi^{AB}\psi^{C}_{\alpha})
+{\bar{\kappa}\over
N}\Tr(\bar{\phi}_{A'B'}\bar{\psi}^{\dot{\alpha}}_{C'})\Tr(\bar{\phi}_{AB}\bar{\psi}_{C\dot{\alpha}})~.~~~\eea
In order to generate operator ${\mathcal{O}}^{[1/2]}_{\I}$, we
should shift $\bar{\phi}_{n+1},\bar{\psi}_{n+2}$. However, there are
two ways of shifting, and their large $z$ behaviors are different.
If we consider $\spab{\bar{\phi}_{n+1}|\bar{\psi}_{n+2}}$-shifting,
the leading term in $z$ is $O(z^0)$, and the boundary operator after
considering the LSZ reduction is
\bea
\mathcal{O}^{\spab{\bar{\phi}_{n+1}|\bar{\psi}_{n+2}}}=\lambda_{n+2,\alpha}\Tr(\phi\psi^{\alpha})~,~~~\eea
hence it has a $\lambda_{n+2,\alpha}$ factor difference with
$\mathcal{O}_{\I}^{[1/2]}$. If we consider
$\spab{\bar{\psi}_{n+2}|\bar{\phi}_{n+1}}$-shifting, the leading
term in $z$ is $O(z)$ order. The boundary operator associated with
the $O(z^0)$ term is quite complicated, but in the $O(z)$ order, we
have
\bea
\mathcal{O}_z^{\spab{\bar{\psi}_{n+2}|\bar{\phi}_{n+1}}}=-\lambda_{n+1,\alpha}\Tr(\phi\psi^{\alpha})~.~~~\eea
These two ways of shifting would give the same result for form
factor of $\mathcal{O}_{\I}^{[1/2]}$. However, it is better to take
the shifting where the leading $z$ term has lower rank, preferably
$O(z^0)$ order, since the computation would be simpler.

The $\Delta L$ term introduces $\phi$-$\psi$-$\phi$-$\psi$ and
$\bar{\phi}$-$\bar{\psi}$-$\bar{\phi}$-$\bar{\psi}$ vertices in the
Feynman diagrams. It is easy to know from Feynman diagram
computation that
$A_{2;2}(\bar{\phi}_1,\bar{\psi}_2;\bar{\phi}_3,\bar{\psi}_4)=\spaa{4~2}$,
and
\bea
A_{3;2}(\bar{\phi}_1,\bar{\psi}_2,g_3^+;\bar{\phi}_4,\bar{\psi}_5)&=&{\spaa{5|P_{23}|\gamma_\mu|2}\over
s_{23}}\epsilon_{3}^{+\mu}-{\spaa{5~2}\over
s_{13}}(p_1-P_{13})_{\mu}\epsilon_{3}^{+\mu}\nonumber\\
&=&{\spaa{1~2}^2\spaa{2~5}\over\spaa{1~2}\spaa{2~3}\spaa{3~1}}~.~~~\eea
This result can be generalized to
\bea
A_{n;2}(\{g^+\},\bar{\phi}_i,\bar{\psi}_{j};\bar{\phi}_{n+1},\bar{\psi}_{n+2})={\spaa{i~j}^2\spaa{j,n+2}\over\spaa{1~2}\spaa{2~3}\cdots\spaa{n~1}}~,~~~\eea
and similarly be proven by BCFW recursion relation. Note that this
amplitude depends on $p_{n+2}$(more strictly speaking,
$\lambda_{n+2}^{\alpha}$) but not $p_{n+1}$, if we take
$\spab{\bar{\phi}_{n+1}|\bar{\psi}_{n+2}}$-shifting, the boundary
contribution equals to the amplitude itself. Thus subtracting the
factor\footnote{We take the convention that
$\spaa{i~j}=\epsilon_{\alpha\beta}\lambda_i^\alpha\lambda_j^\beta=\lambda_i^\alpha\lambda_{j\alpha}$,
$\spbb{i~j}=\epsilon^{\dot{\alpha}\dot{\beta}}\widetilde{\lambda}_{i\dot{\alpha}}\widetilde{\lambda}_{j\dot{\beta}}=\widetilde{\lambda}_{i\dot{\alpha}}\widetilde{\lambda}_{j}^{\dot{\alpha}}$.}
$\lambda_{n+2,\alpha}$, we obtain the form factor of operator
$\mathcal{O}_{\I}^{[1/2]}$ as
\bea
\boxed{\mathcal{F}^{\alpha}_{\mathcal{O}^{[1/2]}_{\I},n}(\{g^+\},\bar{\phi}_i,\bar{\psi}_j;q)={\spaa{i~j}^2\lambda_j^\alpha\over
\spaa{1~2}\spaa{2~3}\cdots\spaa{n~1}}~.~~~}\label{formfactor12I}\eea
If we instead take
$\spab{\bar{\psi}_{n+2}|\bar{\phi}_{n+1}}$-shifting, the boundary
contribution of amplitude $A_{n;2}$ is
\bea
B^{\spab{\bar{\psi}_{n+2}|\bar{\phi}_{n+1}}}_{n;2}(\{g^+\},\bar{\phi}_i,\bar{\psi}_{j};\bar{\phi}_{n+1},\bar{\psi}_{n+2})={\spaa{i~j}^2(\spaa{j,n+2}-z\spaa{j,n+1})\over\spaa{1~2}\spaa{2~3}\cdots\spaa{n~1}}~.~~~\eea
The coefficient of $z$ in above result is identical to the form
factor of
$\mathcal{O}_z^{\spab{\bar{\psi}_{n+2}|\bar{\phi}_{n+1}}}$, and in
order to get the form factor of $\mathcal{O}_{\I}^{[1/2]}$, we
should subtract $-\lambda_{n+1,\alpha}$. The final result is again
(\ref{formfactor12I}).
\begin{figure}
\centering
  \includegraphics[width=6in]{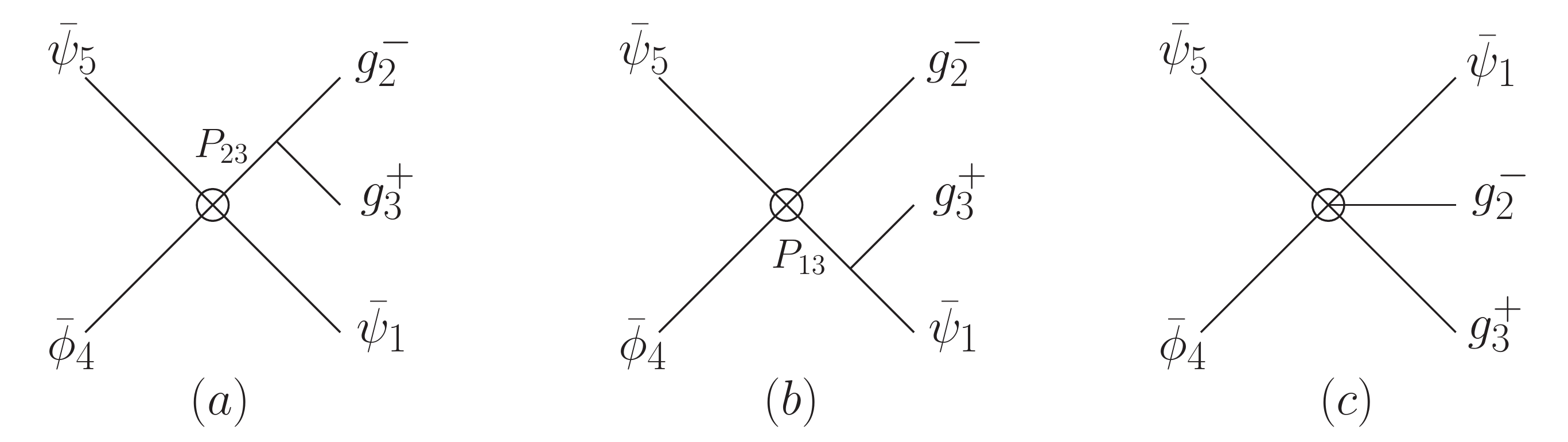}\\
  \caption{Feynman diagrams for $A_{3;2}(\bar{\psi}_1,g_2^-,g_3^+;\bar{\phi}_4,\bar{\psi}_5)$
  defined by $L_{\mathcal{O}^{[1/2]}_{\II}}$. All external particles are out-going.}\label{spsipsigg}
\end{figure}

For operator $\mathcal{O}_{\II}^{[1/2]}$, we can construct the
Lagrangian as
\bea L_{\mathcal{O}_{\II}^{[1/2]}}=L_{\SYM}+{\kappa\over
N}\Tr(\phi^{A'B'}\psi^{C'}_{\alpha})\Tr(\psi^A_{\beta}F^{\beta\alpha})
+{\bar{\kappa}\over
N}\Tr(\bar{\phi}_{AB}\bar{\psi}_{C\dot{\alpha}})\Tr(\bar{\psi}_{A\dot{\beta}}\bar{F}^{\dot{\beta}\dot{\alpha}})~.~~~\eea
Here we choose $\spab{\bar{\phi}_{n+1}|\bar{\psi}_{n+2}}$-shifting
so that the leading term in $z$ is $O(z^0)$ order. The corresponding
boundary operator is
\bea
\mathcal{O}^{\spab{\bar{\phi}_{n+1}|\bar{\psi}_{n+2}}}=\lambda_{n+2,\alpha}\Tr(\psi_{\beta}F^{\beta\alpha})~.~~~\eea
The $\Delta L$ term introduces
four-point(scalar-fermion-fermion-gluon) and
five-point(scalar-fermion-fermion-gluon-gluon) vertices. The
four-point amplitude defined by the four-point vertex is given by
\bea
A_{2;2}(\bar{\psi}_1,g_2^-;\bar{\phi}_3,\bar{\psi}_4)&=&{\spaa{1|2|\gamma_{\mu}|4}+\spaa{4|2|\gamma_{\mu}|1}\over
2}\epsilon_2^{-\mu}=\spaa{1~2}\spaa{4~2}~.~~~\eea
The five-point amplitude
$A_{3;2}(\bar{\psi}_1,g_2^-,g_3^+;\bar{\phi}_4,\bar{\psi}_5)$ can be
computed from three Feynman diagrams as shown in Figure
(\ref{spsipsigg}). The first diagram gives
\bea (a)&=&{1\over 2}\Big(-{\spaa{1|P_{23}|\gamma_{\mu}|5}\over
s_{23}}-{\spaa{5|P_{23}|\gamma_{\mu}|1}\over
s_{23}}\Big)\Big(-(P_{23}\cdot\epsilon_3^+)\epsilon_2^{-\mu}+(p_3\cdot\epsilon_2^{-\mu})\epsilon_3^{+\mu}+(\epsilon_3^+\cdot\epsilon_2^-)p_2^{\mu}\Big)\nonumber\\
&=&{\spaa{2~r_3}\spaa{1~2}\spaa{2~5}\over
\spaa{2~3}\spaa{r_3~3}}-{1\over
2}{\spaa{1~2}\spbb{r_2~3}\spaa{r_3~5}\over
\spaa{r_3~3}\spbb{2~r_2}}-{1\over
2}{\spaa{5~2}\spbb{r_2~3}\spaa{r_3~1}\over
\spaa{r_3~3}\spbb{2~r_2}}~,~~~\eea
and the second diagram gives
\bea (b)&=&-{\spaa{5~2}\spaa{2|P_{13}|\gamma_\mu|1}\over
s_{13}}\epsilon_{3}^{+\mu}={\spaa{2~5}\spaa{1~2}\spaa{r_3~1}\over
\spaa{1~3}\spaa{r_3~3}}~,~~~\eea
while the third diagram gives
\bea
(c)&=&{\spaa{1|\gamma_\mu\gamma_\nu|5}+\spaa{5|\gamma_\mu\gamma_\nu|1}\over
2}\epsilon_2^{-\mu}\epsilon_3^{\nu} ={1\over
2}{\spaa{1~2}\spbb{r_2~3}\spaa{r_3~5}\over
\spaa{r_3~3}\spbb{2~r_2}}+{1\over
2}{\spaa{5~2}\spbb{r_2~3}\spaa{r_3~1}\over
\spaa{r_3~3}\spbb{2~r_2}}~.~~~\eea
Summing above contributions, we get
\bea
A_{3;2}(\bar{\psi}_1,g_2^-,g_3^+;\bar{\phi}_4,\bar{\psi}_5)={\spaa{1~2}^3\spaa{2~5}\over
\spaa{1~2}\spaa{2~3}\spaa{3~1}}~.~~~\eea
Then it is simple to generalize it to
\bea
A_{n;2}(\{g^+\},\bar{\psi}_i,g_j^-;\bar{\phi}_{n+1},\bar{\psi}_{n+2})={\spaa{i~j}^3\spaa{j,n+2}\over
\spaa{1~2}\spaa{2~3}\dots\spaa{n~1}}~,~~~\eea
which can be proven by BCFW recursion relation. Taking
$\spab{\bar{\phi}_{n+1}|\bar{\psi}_{n+2}}$-shifting and Subtracting
$\lambda_{n+2,\alpha}$, we get the form factor
\bea
\boxed{\mathcal{F}^{\alpha}_{\mathcal{O}^{[1/2]}_{\II},n}(\{g^+\},\bar{\psi}_i,g_j^-;q)={\spaa{i~j}^3\lambda_j^{\alpha}\over
\spaa{1~2}\spaa{2~3}\dots\spaa{n~1}}~.~~~}\eea
%

%%%%%%%%%%%%%%%%%%%%%%%%%%%%%
\subsection{The spin-1 operators}
%%%%%%%%%%%%%%%%%%%%%%%%%%%%%

There are three spin-1 operators
\bea
\mathcal{O}^{[1]}_{\I}=\Tr(\psi^{A\alpha}\psi^{B\beta}+\psi^{A\beta}\psi^{B\alpha})~~~,~~~\mathcal{O}^{[1]}_{\II}=\Tr(\phi^{AB}F^{\alpha\beta})
~~~,~~~\mathcal{O}^{[1]}_{\III}=\Tr(\psi^{A\alpha}\bar{\psi}_B^{\dot{\alpha}})~,~~~\eea
and their complex conjugates. In order to construct the Lagrangian,
we need to product them with spin-1 trace term. Since a computation
involving $F^{\alpha\beta}$ is always harder than those involving
fermion and scalar, it is better to choose the trace of two
fermions.

For operator $\mathcal{O}_{\I}^{[1]}$, we can construct the
Lagrangian as
\bea L_{\mathcal{O}_{\I}^{[1]}}=L_{\SYM}+({\kappa \over
N}\Tr(\psi^{A'}_{\alpha}\psi^{B'}_{\beta}+\psi^{A'}_{\beta}\psi^{B'}_{\alpha})\Tr(\psi^{A\alpha}\psi^{B\beta}+\psi^{A\beta}\psi^{B\alpha})+c.c.)~.~~~\eea
Here in order to generate operator $\mathcal{O}_{\I}^{[1]}$, we
should shift two fermions $\bar{\psi}_{n+1},\bar{\psi}_{n+2}$.
Taking $\spab{\bar{\psi}_{n+1}|\bar{\psi}_{n+2}}$-shifting and
considering the LSZ reduction, we find that the leading term in $z$
is $O(z)$ order, and the corresponding boundary operator is
\bea
\mathcal{O}_{z}^{\spab{\bar{\psi}_{n+1}|\bar{\psi}_{n+2}}}=-2\lambda_{n+2,\alpha}\lambda_{n+2,\beta}\Tr(\psi^{A\alpha}\psi^{B\beta}+\psi^{A\beta}\psi^{B\alpha})~.~~~\eea
Thus we also need to take the $O(z)$ order term in the boundary
contribution of amplitude $A_{n;2}$ under
$\spab{\bar{\psi}_{n+1}|\bar{\psi}_{n+2}}$-shifting.

The $\Delta L$ Lagrangian term introduces four-fermion vertex, which
defines the four-point amplitude
$A_{2;2}(\bar{\psi}_1,\bar{\psi}_2;\bar{\psi}_3,\bar{\psi}_4)=\spaa{3~1}\spaa{2~4}+\spaa{4~1}\spaa{2~3}$.
For five-point amplitude
$A_{3;2}(\bar{\psi}_1,\bar{\psi}_2,g_3^+;\bar{\psi}_4,\bar{\psi}_5)$,
there are two contributing Feynman diagrams, and the first diagram
gives
\bea (a)&=&-\spaa{5~2}{\spaa{1|\gamma_\mu|P_{13}|4}\over
s_{13}}\epsilon_3^{+\mu}-\spaa{4~2}{\spaa{1|\gamma_\mu|P_{13}|5}\over
s_{13}}\epsilon_3^{+\mu}\nonumber\\
&=&-{\spaa{5~2}\spaa{4~1}\spaa{1~r_3}\over
\spaa{3~1}\spaa{r_3~3}}-{\spaa{4~2}\spaa{5~1}\spaa{1~r_3}\over
\spaa{3~1}\spaa{r_3~3}}~,~~~\eea
while the second gives
\bea (b)&=&\spaa{5~1}{\spaa{2|\gamma_{\mu}|P_{23}|4}\over
s_{23}}\epsilon_3^{+\mu}+\spaa{4~1}{\spaa{2|\gamma_{\mu}|P_{23}|5}\over
s_{23}}\epsilon_3^{+\mu}\nonumber\\
&=&{\spaa{5~1}\spaa{4~2}\spaa{2~r_3}\over
\spaa{3~2}\spaa{r_3~3}}+{\spaa{4~1}\spaa{5~2}\spaa{2~r_3}\over
\spaa{3~2}\spaa{r_3~3}}~.~~~\eea
Thus
\bea
A_{3;2}(\bar{\psi}_1,\bar{\psi}_2,g_3^+;\bar{\psi}_4,\bar{\psi}_5)&=&{\spaa{1~2}^2\over\spaa{1~2}\spaa{2~3}\spaa{3~1}}(\spaa{4~1}\spaa{2~5}+\spaa{5~1}\spaa{2~4})~.~~~\eea
By BCFW recursion relation, we also have
\bea
&&A_{n;2}(\{g^+\},\bar{\psi}_i,\bar{\psi}_j;\bar{\psi}_{n+1},\bar{\psi}_{n+2})\nonumber\\
&&={\spaa{i~j}^2\over
\spaa{1~2}\spaa{2~3}\cdots\spaa{n~1}}(\spaa{n+1,i}\spaa{j,n+2}+\spaa{n+2,i}\spaa{j,n+1})~.~~~\eea
Notice that this amplitude depends on both
$\lambda_{n+1}^{\alpha},\lambda_{n+2}^{\alpha}$, thus the $O(z)$
term is unavoidable when shifting two fermions. The boundary
contribution under
$\spab{\bar{\psi}_{n+1}|\bar{\psi}_{n+2}}$-shifting is
\bea
&&B_{n;2}^{\spab{\bar{\psi}_{n+1}|\bar{\psi}_{n+2}}}(\{g^+\},\bar{\psi}_i,\bar{\psi}_j;\bar{\psi}_{\widehat{n+1}},\bar{\psi}_{\widehat{n+2}})\nonumber\\
&&=-2z{\spaa{i~j}^2\over
\spaa{1~2}\spaa{2~3}\cdots\spaa{n~1}}\spaa{n+2,i}\spaa{j,n+2}\nonumber\\
&&~~~~~~~+{\spaa{i~j}^2\over
\spaa{1~2}\spaa{2~3}\cdots\spaa{n~1}}(\spaa{n+1,i}\spaa{j,n+2}+\spaa{n+2,i}\spaa{j,n+1})~.~~~\eea
Taking the $O(z)$ contribution and subtracting the factor
$-2\lambda_{n+2,\alpha}\lambda_{n+2,\beta}$, we get the form factor
\bea
\boxed{\mathcal{F}^{\alpha\beta}_{\mathcal{O}_1^{[1]},n}(\{g^+\},\bar{\psi}_i,\bar{\psi}_j;q)={\spaa{i~j}^2\over
\spaa{1~2}\spaa{2~3}\cdots\spaa{n~1}}\left({\lambda_i^{\alpha}\lambda_j^{\beta}+\lambda_j^{\alpha}\lambda_i^{\beta}\over
2}\right)~,~~~}\eea
where we have symmetrized the indices $\alpha,\beta$.

\begin{figure}
\centering
  \includegraphics[width=6in]{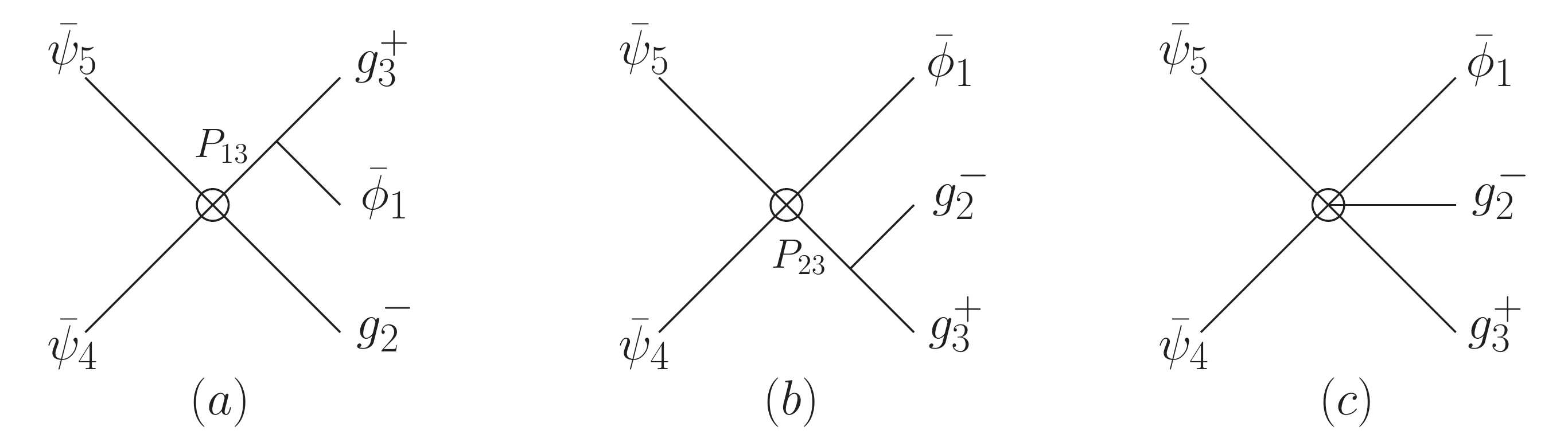}\\
  \caption{Feynman diagrams for $A_{3;2}(\bar{\phi}_1,g_2^-,g_3^+;\bar{\psi}_4,\bar{\psi}_5)$
  defined by $L_{\mathcal{O}^{[1]}_{\II}}$. All external particles are out-going.}\label{psipsisgg}
\end{figure}

Similar construction can be applied to the operator
$\mathcal{O}_{\II}^{[1]}$, where we have
\bea L_{\mathcal{O}_{\II}^{[1]}}=L_{\SYM}+({\kappa\over
N}\Tr(\psi^{A'}_{\alpha}\psi^{B'}_{\beta}+\psi^{A'}_{\beta}\psi^{B'}_{\alpha})\Tr(\phi^{AB}F^{\alpha\beta})+c.c.)~.~~~\eea
The leading term in $z$ under
$\spab{\bar{\psi}_{n+1}|\bar{\psi}_{n+2}}$-shifting is $O(z)$ order,
and the boundary operator is
\bea
\mathcal{O}_z^{\spab{\bar{\psi}_{n+1}|\bar{\psi}_{n+2}}}=-\lambda_{n+2,\alpha}\lambda_{n+2,\beta}\Tr(\phi^{AB}F^{\alpha\beta})~.~~~
\eea
The $\Delta L$ Lagrangian term introduces
four-point(fermion-fermion-scalar-gluon) vertex and
five-point(fermion-fermion-scalar-gluon-gluon) vertex. The
four-point vertex defines four-point amplitude
$A_{2;2}(\bar{\phi}_1,g_2^-;\bar{\psi}_3,\bar{\psi}_4)=-{1\over
2}(\spaa{3|2|\gamma_{\mu}|4}+\spaa{4|2|\gamma_\mu|3})\epsilon_2^{-\mu}=\spaa{2~3}\spaa{2~4}$,
while for five-point amplitude
$A_{3;2}(\bar{\phi}_1,g_2^-,g_3^+;\bar{\psi}_4,\bar{\psi}_5)$, we
need to consider three Feynman diagrams, as shown in Figure
(\ref{psipsisgg}).
The first diagram gives
\bea (a)&=&{\spaa{2~4}\spaa{2~5}\over
s_{13}}(p_1+P_{13})_{\mu}\epsilon_{3}^{+\mu}={\spaa{2~4}\spaa{2~5}\spaa{r_3~1}\over
\spaa{3~1}\spaa{r_3~3}}~,~~~\eea
the second diagram gives
\bea (b)&=&{1\over 2}\Big(-{\spaa{4|P_{23}|\gamma_{\mu}|5}\over
s_{23}}-{\spaa{5|P_{23}|\gamma_{\mu}|4}\over
s_{23}}\Big)\Big(-(P_{23}\cdot\epsilon_3^+)\epsilon_2^{-\mu}+(p_3\cdot\epsilon_2^-)\epsilon_3^{\mu}+(\epsilon_3^+\cdot\epsilon_2^{-})p_2^{\mu}\Big)\nonumber\\
&=&{\spaa{r_3~2}\spaa{2~4}\spaa{2~5}\over
\spaa{2~3}\spaa{r_3~3}}+{1\over
2}{\spbb{r_2~3}\spaa{r_3~4}\spaa{2~5}\over
\spaa{r_3~3}\spbb{2~r_2}}+{1\over
2}{\spbb{r_2~3}\spaa{r_3~5}\spaa{2~4}\over
\spaa{r_3~3}\spbb{2~r_2}}~,~~~\eea
and the third diagram gives
\bea
(c)&=&{\spaa{4|\gamma_{\mu}\gamma_{\nu}|5}+\spaa{5|\gamma_{\mu}\gamma_{\nu}|4}\over
2}\epsilon_2^{-\mu}\epsilon_3^{+\nu}={1\over
2}{\spaa{4~2}\spbb{r_2~3}\spaa{r_3~5}\over
\spaa{r_3~3}\spbb{2~r_2}}+{1\over
2}{\spaa{5~2}\spbb{r_2~3}\spaa{r_3~4}\over
\spaa{r_3~3}\spbb{2~r_2}}~.~~~\eea
Summing above contributions, we get
\bea
A_{3;2}(\bar{\phi}_1,g_2^-,g_3^+;\bar{\psi}_4,\bar{\psi}_5)&=&{\spaa{1~2}^2\over\spaa{1~2}\spaa{2~3}\spaa{3~1}}\spaa{4~2}\spaa{2~5}~.~~~\eea
Generalizing above result to $(n+2)$-point amplitude, we have
\bea
A_{n;2}(\{g^+\},\bar{\phi}_i,g_j^-;\bar{\psi}_{n+1},\bar{\psi}_{n+2})={\spaa{i~j}^2\over
\spaa{1~2}\spaa{2~3}\cdots\spaa{n~1}}\spaa{n+1,j}\spaa{j,n+2}~,~~~\eea
which can be trivially proven by BCFW recursion relation. We are
only interested in the $O(z)$ term of the boundary contribution
under $\spab{\bar{\psi}_{n+1}|\bar{\psi}_{n+2}}$-shifting, which is
\bea
B_{n;2}^{\spab{\bar{\psi}_{n+1}|\bar{\psi}_{n+2}}}(\{g^+\},\bar{\phi}_i,g_j^-;\bar{\psi}_{\widehat{n+1}},\bar{\psi}_{\widehat{n+2}})=-z{\spaa{i~j}^2\spaa{n+2,j}\spaa{j,n+2}\over
\spaa{1~2}\spaa{2~3}\cdots\spaa{n~1}}+O(z^0)~.~~~\eea
After subtracting the factor
$-\lambda_{n+2,\alpha}\lambda_{n+2,\beta}$, we get
\bea
\boxed{\mathcal{F}^{\alpha\beta}_{\mathcal{O}_{\II}^{[0]},n}(\{g^+\},\bar{\phi}_i,g_j^-;q)={\spaa{i~j}^2\over
\spaa{1~2}\spaa{2~3}\cdots\spaa{n~1}}(-\lambda_j^{\alpha}\lambda_j^{\beta})~.~~~}\eea

Now let us turn to the operator $\mathcal{O}_{\III}^{[1]}$, and
construct the Lagrangian as
\bea L_{\mathcal{O}_{\III}^{[1]}}=L_{\SYM}+{\kappa\over
N}\Tr(\psi^{A'}_{\alpha}\bar{\psi}_{B'\dot{\alpha}})\Tr(\psi^{A\alpha}\bar{\psi}_{B}^{\dot{\alpha}})~.~~~\eea
The leading term in $z$ under
$\spab{\bar{\psi}_{n+2}|\psi_{n+1}}$-shifting is $O(z^2)$ order,
while the leading term in $z$ under
$\spab{\psi_{n+1}|\bar{\psi}_{n+2}}$-shifting is $O(z^0)$ order. In
the later case, the boundary operator is
\bea
\mathcal{O}^{\spab{\psi_{n+1}|\bar{\psi}_{n+2}}}=\widetilde{\lambda}_{n+1,\dot{\alpha}}\lambda_{n+2,\alpha}\Tr(\psi^{A\alpha}\bar{\psi}^{\dot{\alpha}}_B)~.~~~\eea
The four-point amplitude
$A_{2;2}(\psi_1,\bar{\psi}_2;\psi_3,\bar{\psi}_4)=\spbb{1~3}\spaa{2~4}$,
while the five-point amplitude
\bea
A_{3;2}(\psi_1,\bar{\psi}_2,g_3^+;\psi_4,\bar{\psi}_5)&=&\spaa{2~5}{\spbb{1|\gamma_\mu|P_{13}|4}\over
s_{13}}\epsilon_3^{+\mu}-\spbb{1~4}{\spaa{2|\gamma_\mu|P_{23}|5}\over
s_{23}}\epsilon_3^{+\mu}\nonumber\\
&=&{\spaa{1~2}\spaa{2~5}\spab{2|1+3|4}\over
\spaa{1~2}\spaa{2~3}\spaa{3~1}}~.~~~\eea
Note that $\spab{2|1+3|4}=\spab{2|1+2+3|4}=\spab{2|q|4}$, where
$q=-p_4-p_5$, we can generalize above result to $(n+2)$-point as
\bea
A_{n;2}(\{g^+\},\psi_i,\bar{\psi}_j;\psi_{n+1},\bar{\psi}_{n+2})={\spaa{i~j}\spaa{j,n+2}\spab{j|q|n+1}\over\spaa{1~2}\spaa{2~3}\cdots\spaa{n~1}}~,~~~\label{formfactor1III}\eea
where $q=-p_{n+1}-p_{n+2}$. Let us verify eqn.
(\ref{formfactor1III}) by induction method. Assuming eqn.
(\ref{formfactor1III}) is valid for $A_{n-1;2}$, and taking
$\spab{g_1^+|g_n^+}$-shifting, we get two contributing
terms\footnote{We have assumed that $i,j\neq 2,n-1$, otherwise the
contributing terms are slightly different. But the conclusion is the
same.} from BCFW expansion. One is
\bea A_{n-1;2}(g^+_{\widehat{1}},\ldots, \psi_i,\ldots,
\bar{\psi}_j,\ldots,
g^+_{n-2},g^+_{\widehat{P}_{n-1,n}};\psi_{n+1},\bar{\psi}_{n+2}){1\over
P^2_{n-1,n}}A_3(g^-_{-\widehat{P}_{n-1,n}},g^+_{n-1},g^+_{\widehat{n}})~.~~~\nonumber\eea
Since $\widehat{P}_{n-1,n}^2=\spaa{n-1,n}\spbb{\widehat{n},n-1}=0$,
so $A_3(g^-_{\widehat{P}_{n-1,n}},g^+_{n-1},g^+_{\widehat{n}})\sim
\spbb{n-1,\widehat{n}}^3\to 0$, and this term vanishes. The other
contributing term is
\bea &&A_{3}(g^+_{\widehat{1}},g^+_2,g^-_{\widehat{P}_{12}}){1\over
P_{12}^2}A_{n-1;2}(g^+_{-\widehat{P}_{12}},g^+_3,\ldots,\psi_i,\ldots,
\bar{\psi}_j,\ldots,
g^+_{\widehat{n}};\psi_{n+1},\bar{\psi}_{n+2})~.~~~\eea
By inserting the explicit expressions of $A_3$ and $A_{n-1;2}$, we
arrive at eqn. (\ref{formfactor1III}).

Under $\spab{\psi_{n+1}|\bar{\psi}_{n+2}}$-shifting, the boundary
contribution is
\bea
B_{n;2}^{\spab{\psi_{n+1}|\bar{\psi}_{n+2}}}(\{g^+\},\psi_i,\bar{\psi}_j;\psi_{\widehat{n+1}},\bar{\psi}_{\widehat{n+2}})={\spaa{i~j}\spaa{j,n+2}\spab{j|q|n+1}\over\spaa{1~2}\spaa{2~3}\cdots\spaa{n~1}}~.~~~\eea
So subtracting the factor
$\lambda_{n+2,\alpha}\widetilde{\lambda}_{n+1,\dot{\alpha}}$, we get
the form factor
\bea
\boxed{\mathcal{F}^{\alpha\dot{\alpha}}_{\mathcal{O}_{\III}^{[1]},n}(\{g^+\},\psi_i,\bar{\psi}_j;q)={\spaa{i~j}\over\spaa{1~2}\spaa{2~3}\cdots\spaa{n~1}}\lambda_j^{\alpha}(\lambda_{j\beta}q^{\beta\dot{\alpha}})~.~~~}\eea
%

%%%%%%%%%%%%%%%%%%%%%%%
\subsection{The spin-${3\over 2}$ operators}
%%%%%%%%%%%%%%%%%%%%%%%%%%%

There are two operators
\bea
\mathcal{O}_{\I}^{[3/2]}=\Tr(\bar{\psi}^{\dot{\alpha}}F^{\alpha\beta})~~~~,~~~~\mathcal{O}_{\II}^{[3/2]}=\Tr(\psi^{\gamma}F^{\alpha\beta})~~~~\eea
with their complex conjugate partners. We need to product them with
spin-${3\over 2}$ trace term to construct $\Delta L$.

For the operator $\mathcal{O}_{\I}^{[3/2]}$, we can construct the
Lagrangian
as
\bea L_{\mathcal{O}_{\I}^{[3/2]}}=L_{\SYM}+{\kappa\over
N}\Tr(\bar{\psi}_{\dot{\alpha}}F_{\alpha\beta})\Tr(\bar{\psi}^{\dot{\alpha}}F^{\alpha\beta})+{\bar{\kappa}\over
N}\Tr(\psi_{\alpha}\bar{F}_{\dot{\alpha}\dot{\beta}})\Tr(\psi^{\alpha}\bar{F}^{\dot{\alpha}\dot{\beta}})~.~~~\eea
It introduces new four-point vertices
$\bar{\psi}$-$g^+$-$\bar{\psi}$-$g^+$ and $\psi$-$g^-$-$\psi$-$g^-$,
as well as five, six-point vertices.

From Feynman diagrams, we can directly compute
$A_{2;2}(\psi_1,g_2^-;\psi_3,g_4^-)=\spaa{2~4}^2\spbb{1~3}$, while
for the five-point amplitude
$A_{3;2}(\psi_1,g_2^-,g_3^+;\psi_4,g_5^-)$, we need to compute three
Feynman diagrams, which are given by
\bea (a)&=&\spaa{2~5}^2{\spbb{1|\gamma_\mu|P_{13}|4}\over
s_{13}}\epsilon_3^{+\mu}={\spaa{2~5}^2\spbb{3~4}\over
\spaa{3~1}}+{\spaa{2~5}^2\spbb{1~4}\spaa{r_3~1}\over
\spaa{3~1}\spaa{r_3~3}}~,~~~\eea
\bea (b)&=&-\spbb{1~4}{\spaa{5|P_{23}|\gamma_\mu|5}\over
s_{23}}\Big(-(P_{23}\cdot\epsilon_3^+)\epsilon_2^{-\mu}+(p_3\cdot\epsilon_2^-)\epsilon_3^{+\mu}+(\epsilon_3^+\cdot\epsilon_2^{-})p_2^{\mu}\Big)\nonumber\\
&=&{\spbb{1~4}\spaa{2~5}^2\spaa{r_3~2}\over
\spaa{2~3}\spaa{r_3~3}}+{\spbb{1~4}\spaa{2~5}\spbb{r_2~3}\spaa{r_3~5}\over
\spaa{r_3~3}\spbb{2~r_2}}~,~~~\eea
and
\bea
(c)=\spbb{1~4}\spaa{5|\gamma_\mu\gamma_\nu|5}\epsilon_2^{-\mu}\epsilon_3^{+\nu}={\spbb{1~4}\spaa{5~2}\spbb{r_2~3}\spaa{r_3~5}\over\spaa{r_3~3}\spaa{2~r_2}}~.~~~\eea
So the final result is
\bea
A_{3;2}(\psi_1,g_2^-,g_3^+;\psi_4,g_5^-)={\spaa{1~2}\spaa{2~5}^2\spab{2|q|4}\over
\spaa{1~2}\spaa{2~3}\spaa{3~1}}~,~~~\eea
where $q=-p_4-p_5$. This result can be generalized to
\bea
A_{n;2}(\{g^+\},\psi_i,g_j^-;\psi_{n+1},g^-_{n+2})={\spaa{i~j}\spaa{j,n+2}^2\spab{j|q|n+1}\over
\spaa{1~2}\spaa{2~3}\cdots\spaa{n~1}}~,~~~\eea
where $q=-p_{n+1}-p_{n+2}$, and proven by BCFW recursion relation as
done for the $\mathcal{O}_{\III}^{[1]}$ case.

If taking $\spab{g_{n+2}^-|\psi_{n+1}}$-shifting, the leading $z$
term in the boundary operator would be $O(z^3)$ order. We can
however choose $\spab{\psi_{n+1}|g_{n+2}^-}$-shifting, under which
there is only $O(z^0)$ term in the boundary operator,
\bea
\mathcal{O}^{\spab{\psi_{n+1}|g_{n+2}^-}}=\widetilde{\lambda}_{n+1,\dot{\alpha}}\lambda_{n+2,\alpha}\lambda_{n+2,\beta}\Tr(\bar{\psi}^{\dot{\alpha}}F^{\alpha\beta})~.~~~\eea
The boundary contribution of amplitude $A_{n;2}$ under
$\spab{\psi_{n+1}|g_{n+2}^-}$-shifting equals to $A_{n;2}$ itself,
thus after subtracting factor
$\widetilde{\lambda}_{n+1,\dot{\alpha}}\lambda_{n+2,\alpha}\lambda_{n+2,\beta}$,
we get the form factor
\bea
\boxed{\mathcal{F}^{\dot{\alpha}~\alpha\beta}_{\mathcal{O}_{\I}^{[3/2]},n}(\{g^+\},\psi_i,g_j^-;q)={\spaa{i~j}\over
\spaa{1~2}\spaa{2~3}\cdots\spaa{n~1}}\lambda_j^{\alpha}\lambda_j^{\beta}(\lambda_{j\gamma}q^{\gamma\dot{\alpha}})~.~~~}\eea

Discussion on the operator $\mathcal{O}^{[3/2]}_{\II}$ is almost the
same as operator $\mathcal{O}^{[3/2]}_{\I}$, while we only need to
change $\psi\to\bar{\psi}$. We can construct the Lagrangian as
\bea L_{\mathcal{O}_{\II}^{[3/2]}}=L_{\SYM}+{\kappa\over
N}\Tr(\psi_{\gamma}F_{\alpha\beta})\Tr(\psi^{\gamma}F^{\alpha\beta})+{\bar{\kappa}\over
N}\Tr(\bar{\psi}_{\dot{\gamma}}\bar{F}_{\dot{\alpha}\dot{\beta}})\Tr(\bar{\psi}^{\dot{\gamma}}\bar{F}^{\dot{\alpha}\dot{\beta}})~.~~~\eea
In order to generate the operator
$\Tr(\psi^{\gamma}F^{\alpha\beta})$, we need to shift
$\bar{\psi}_{n+1}, g^-_{n+2}$. Under
$\spab{g_{n+2}^-|\bar{\psi}_{n+1}}$-shifting, the leading term in
$z$ is $O(z^2)$ order, and the corresponding boundary operator is
\bea
\mathcal{O}^{\spab{g_{n+2}^-|\bar{\psi}_{n+1}}}_{z^2}=\lambda_{n+1,\alpha}\lambda_{n+1,\beta}\lambda_{n+1,\gamma}\Tr(\psi^{\gamma}F^{\alpha\beta})~.~~~\eea
We can also take $\spab{\bar{\psi}_{n+1}|g_{n+2}^-}$-shifting, and
the corresponding boundary operator is $O(z)$ order,
\bea
\mathcal{O}_{z}^{\spab{\bar{\psi}_{n+1}|g_{n+2}^-}}=\lambda_{n+2,\gamma}\lambda_{n+2,\alpha}\lambda_{n+2,\beta}\Tr(\psi^{\gamma}F^{\alpha\beta})~.~~~\eea

Computation of double trace amplitudes defined by
$L_{\mathcal{O}_{\II}^{[3/2]}}$ is similar to those defined by
$L_{\mathcal{O}_{\I}^{[3/2]}}$, and we immediately get
$A_{2;2}(\bar{\psi}_1,g_2^-;\bar{\psi}_3,g_4^-)=\spaa{2~4}^2\spaa{3~1}$,
and
\bea
A_{3;2}(\bar{\psi}_1,g_2^-,g_3^+;\bar{\psi}_4,g_5^-)={\spaa{1~2}^2\spaa{2~5}^2\spaa{1~4}\over
\spaa{1~2}\spaa{2~3}\spaa{3~1}}~.~~~\eea
For general $(n+2)$-point amplitude, we have
\bea
A_{n;2}(\{g^+\},\bar{\psi}_i,g_j^-;\bar{\psi}_{n+1},g_{n+2}^-)={\spaa{i~j}^2\spaa{j,n+2}^2\spaa{i,n+1}\over
\spaa{1~2}\spaa{2~3}\cdots\spaa{n~1}}~.~~~\eea
We can either take $\spab{g_{n+2}^-|\bar{\psi}_{n+1}}$-shifting or
$\spab{\bar{\psi}_{n+1}|g_{n+2}^-}$-shifting to compute the form
factor of $\mathcal{O}_{\II}^{[3/2]}$. For example, under
$\spab{g_{n+2}^-|\bar{\psi}_{n+1}}$-shifting, we pick up the
$O(z^2)$ term of boundary contribution, which is
$$z^2{\spaa{i~j}^2\spaa{j,n+1}^2\spaa{i,n+1}\over
\spaa{1~2}\spaa{2~3}\cdots\spaa{n~1}}~,$$ subtract the factor $\lambda_{n+1,\alpha}\lambda_{n+1,\beta}\lambda_{n+1,\gamma}$, and finally get the form factor,
\bea
\boxed{\mathcal{F}^{\alpha\beta\gamma}_{\mathcal{O}_{\II}^{[3/2]},n}(\{g^+\},\bar{\psi}_i,g_j^-;\bar{\psi}_{n+1},g_{n+2}^-)={\spaa{i~j}^2\over\spaa{1~2}\spaa{2~3}\cdots\spaa{n~1}}\lambda_{j}^{\alpha}\lambda_{j}^{\beta}\lambda_{i}^{\gamma}~.~~~}\eea
%

%%%%%%%%%%%%%%%%%%%%%%%%%
\subsection{The spin-2 operator}
\label{sectionspin2}
%%%%%%%%%%%%%%%%%

For the spin-2 operator
\bea
\mathcal{O}_{\I}^{[2]}=\Tr(F^{\alpha\beta}\bar{F}^{\dot{\alpha}\dot{\beta}})~,~~~\eea
we can construct the Lagrangian as
\bea L_{\mathcal{O}_{\I}^{[2]}}=L_{\SYM}+{\kappa\over
N}\Tr(F_{\alpha\beta}\bar{F}_{\dot{\alpha}\dot{\beta}})\Tr(F^{\alpha\beta}\bar{F}^{\dot{\alpha}\dot{\beta}})~.~~~\eea
The $\Delta L$ Lagrangian term introduces four to eight-point gluon
vertices in Feynman diagrams. It is easy to know that the four-point
amplitude
$A_{2;2}(g_1^-,g_2^+;g_3^-,g_4^+)=\spaa{1~3}^2\spbb{2~4}^2$. The
general $(n+2)$-point amplitude is given by
\bea
A_{n;2}(\{g^+\},g_i^-;g_{n+1}^-,g_{n+2}^+)=-{\spab{i|q|n+2}^2\spaa{i,n+1}^2\over\spaa{1~2}\spaa{2~3}\cdots\spaa{n~1}}~,~~~\label{formfactor2I}\eea
where $q=-p_{n+1}-p_{n+2}$. Let us verify this result by BCFW
recursion relation. Assuming eqn. (\ref{formfactor2I}) is valid for
$A_{n-1;2}$, and taking $\spab{g_{n-1}^+|g_{n}^+}$-shifting, we get
two contributing terms\footnote{We have assumed $i\neq 1,n-2$, which
can always be true by cyclic invariance of the external legs.} in
BCFW expansion. The first term is
\bea A_{n-1;2}(g_2^+,\ldots, g_{i-1}^+,g_i^-,g_{i+1}^+,\ldots,
g_{\widehat{n-1}}^+,g_{\widehat{P}_{1n}}^+;g_{n+1}^-,g_{n+2}^+){1\over
P^2_{1n}}A_{3}(g_{-\widehat{P}_{1n}}^-,g_{\widehat{n}}^+,g_1^+)~,~~~\eea
and this one vanishes, since the on-shell condition of propagator
$\widehat{P}_{1n}^2=\spaa{1~n}\spbb{\widehat{n}~1}=0$ implies
$A_{3}(g^-_{-\widehat{P}_{1n}},g_{\widehat{n}}^+,g_1^+)\sim
\spbb{\widehat{n}~1}^3\to 0$. The other term is
%The other one
%is($\widehat{P}_{n-2,n-1}=p_{n-2}+p_{n-1}-z|n\rangle|n-1]$,
%$z_{n-2,n-1}={\spaa{n-2,n-1}\over \spaa{n-2,n}}$)
%
\bea &&A_3(g_{n-2}^+,g_{\widehat{n-1}}^+,g_{\widehat{P}}^-){1\over
P_{n-2,n-1}^2}A_{n-1;2}(g^+_{-\widehat{P}},g_{\widehat{n}}^+,g_1^+,\ldots,g_{i-1}^+,g_i^-,g_{i+1}^+,\ldots,g_{n-3}^+;g_{n+1}^-,g_{n+2}^+)~.~~~\nonumber\eea
After inserting the explicit expressions for $A_{3}$ and
$A_{n-1;2}$, we arrive at the result (\ref{formfactor2I}).

The leading $z$ term of boundary operator under
$\spab{g_{n+1}^-|g_{n+2}^+}$-shifting is $O(z^4)$ order. Instead, we
would like to take $\spab{g_{n+2}^+|g_{n+1}^-}$-shifting, under
which the boundary operator is $O(z^0)$ order. After considering LSZ
reduction, we have
\bea
\mathcal{O}^{\spab{g_{n+2}^+|g_{n+1}^-}}=\widetilde{\lambda}_{n+2,\dot{\alpha}}\widetilde{\lambda}_{n+2,\dot{\beta}}\lambda_{n+1,\alpha}\lambda_{n+1,\beta}\Tr(F^{\alpha\beta}\bar{F}^{\dot{\alpha}\dot{\beta}})~.~~~\eea
Hence by picking up the boundary contribution of amplitude $A_{n;2}$
under $\spab{g_{n+2}^+|g_{n+1}^-}$-shifting, and subtracting factor
$\widetilde{\lambda}_{n+2,\dot{\alpha}}\widetilde{\lambda}_{n+2,\dot{\beta}}\lambda_{n+1,\alpha}\lambda_{n+1,\beta}$,
we get the form factor
\bea
\boxed{\mathcal{F}_{\mathcal{O}_{\I}^{[2]},n}^{\dot{\alpha}\dot{\beta}~\alpha\beta}(\{g^+\},g_i^-;q)=-{(\lambda_{i\gamma_1}q^{\gamma_1\dot{\alpha}})(\lambda_{i\gamma_2}q^{\gamma_2\dot{\beta}})\lambda^{\alpha}_{i}\lambda^{\beta}_{i}\over\spaa{1~2}\spaa{2~3}\cdots\spaa{n~1}}~.~~~}\eea
%

%%%%%%%%%%%%%%%%%
\section{Summary and discussion}
\label{secConclusion}
%%%%%%%%%%%%%%%%

The boundary operator is initially introduced as a formal technique
to study the boundary contribution of amplitude when doing BCFW
recursion relation in paper \cite{Jin:2015pua}. It defines a form
factor, and practically this off-shell quantity is difficult to
compute. In this paper, we take the reversed way to study the form
factor from boundary contribution of amplitude of certain theory. We
show that by suitable construction of Lagrangian, it is possible to
generate boundary operators which are identical(or proportional) to
the given operators of interest. This means that the form factor of
given operator can be extracted from the boundary contribution of
corresponding amplitude defined by that Lagrangian. We demonstrate
this procedure for a class of composite operators by computing
amplitudes of double trace structure and reading out the form
factors from corresponding boundary contribution. Thus the
computation of form factor becomes a problem of computing the
scattering amplitude.

We have considered a class of composite operators, which are traces
of product of two component fields from $\mathcal{N}=4$ SYM, and the
sum of spins of those two fields is no larger than two. In fact, the
construction of Lagrangian has no difference for other operators
with length(the number of fields inside the trace) larger than two,
provided the sum of their spins is no larger than two. This is
because we can always product them with a length-two trace term to
make a Lorentz-invariant Lagrangian term, and deform the two fields
in the extra trace term to produce the required boundary operators.
However, if the operator has spin larger than two, in order to make
a Lorentz-invariant Lagrangian term, the length of extra trace term
should be larger than two. Then deformation of two fields in the
extra trace term is not sufficient to produce the desired boundary
operators, and we need multi-step deformation. It would be
interesting to investigate how this multi-step deformation works
out. It would also be interesting to find out how to apply this
story to other kind of operators such as stress-tensor multiplet or
amplitude with off-shell currents.

Note that all the discussions considered in this paper are at
tree-level. While it is argued\cite{Jin:2015pua} that the boundary
operator is generalizable to loop-level since the OPE can be defined
therein, it is interesting to see if similar connection between form
factor and amplitude also exists at loop-level or not. For this
purpose, it would be better to study the loop corrections to the
boundary operators, which is under investigation.

%%%%%%%%%%%%%%%%%%%
\acknowledgments
%%%%%%%%%%%%%%%%
This work is supported by the Qiu-Shi Fund and the National Natural
Science Foundation of China(Chinese NSF) with Grant No.11135006,
No.11125523, No.11575156. RH would also like to acknowledge the
supporting from Chinese Postdoctoral Administrative Committee.

%This work is supported by Qiu-Shi funding and Chinese NSF funding
%under contracts No.11031005, No.11135006, No.11125523 and
%No.10875103, and National Basic Research Program of China
%(2010CB833000).

%%%%%%%%%%%%%%%%%%%
\appendix
%%%%%%%%%%%%%%%%%%%%%

%%%%%%%%%%%%%%%%%%%%%%
\section{Brief review on constructing the boundary operator}
%%%%%%%%%%%%%%%%%%%%%%%%%

For reader's convenience we briefly review the results of paper
\cite{Jin:2015pua} in this appendix. Please refer to that paper for
more details.

The whole idea is to consider the OPE expansion in momentum space in
the large $z$ limits, and work out the expansion coefficients of
each $z$ order. Denoting the two shifted fields as
$\Phi_{1}^{\Lambda}\equiv \Phi^{\Lambda}(p_1+zq)$ and
$\Phi_{n}^{\Lambda}\equiv \Phi^{\Lambda}(p_n-zq)$, one found that
the $z$-dependence can be computed from
\bea \mathcal{Z}(z)=-i\int
D\Phi^{\Lambda}\exp\Big(iS_2^{\Lambda}[\Phi^\Lambda,\Phi]\Big)\Phi_1^{\Lambda}\Phi_n^{\Lambda}~,~~~\label{Zz}\eea
where $S_{2}^{\Lambda}[\Phi^\Lambda,\Phi]$ is the quadratic term of
$\Phi^{\Lambda}$ in action $S$ after field splitting $\Phi\to
\Phi+\Phi^\Lambda$(soft part and hard part). This can be interpreted
as the OPE of $\Phi_1^{\Lambda}$ and $\Phi_n^{\Lambda}$. Expanding
$\mathcal{Z}(z)$ around $z=\infty$ yields
\bea \mathcal{Z}(z)=\cdots+{1\over
z}\mathcal{O}_{z^{-1}}+\mathcal{O}^{\spab{\phi_1|\Phi_n}}+z\mathcal{O}_z^{\spab{\Phi_1|\Phi_n}}+\cdots~.~~~\eea
In order to construct the boundary operator for given $z$ order, one
should compute $\mathcal{Z}(z)$, i.e., evaluate the integral
(\ref{Zz}). Since $S_2^{\Lambda}$ only contains terms quadratic in
$\Phi^{\Lambda}$, integral (\ref{Zz}) can be evaluated exactly.
Assume a theory has $M$ real fields $\psi^I$ and $N$ complex fields
$\phi^A$, compactly expressed as
\bea \Phi^{\alpha}=\left(
                     \begin{array}{c}
                       \varphi^I \\
                       \phi^A \\
                       \bar{\phi}^A \\
                     \end{array}
                   \right)~~~,~~~H^{\alpha}=\left(
                     \begin{array}{c}
                       \widehat{\varphi}^I \\
                       \widehat{\phi}^A \\
                       \widehat{\bar{\phi}}^A \\
                     \end{array}
                   \right)~,~~~\label{defPhi}\eea
where we have combined hard fields into $H^\alpha$. The complex
conjugates of $\Phi^\alpha$ is
$\Phi^{\dagger}_\alpha=\big(\varphi^I~~\bar{\phi}_A~~\phi^A\big)$,
and be related to $\Phi^\alpha$ as
$\Phi^\alpha=T^{\alpha\beta}\Phi_{\beta}^\dagger$ through matrix
\bea T^{\alpha\beta}=\left(
                       \begin{array}{ccc}
                         I_M & 0 & 0 \\
                         0 & 0 & I_N \\
                         0 & I_N & 0 \\
                       \end{array}
                     \right)~.~~~\label{Phi2DaggerPhi}\eea
With these notations, the quadratic term in the Lagrangian is
\bea L_2^{\Lambda}={1\over 2}H_{\alpha}^{\dagger}
\mathcal{D}^{\alpha}_{~\beta}H^{\beta}~~~,~~~\mathcal{D}^{\alpha}_{~\beta}={\delta^2\over
\delta\Phi^{\dagger}_{\alpha}\delta\Phi^{\beta}}L~.~~~\eea
Following the standard procedure of computing generating functions,
one can get
\bea
\mathcal{Z}(z)=\mathcal{Z}^{\Lambda}[\Phi](\mathcal{D}^{-1})^{\alpha\beta}(x,y;\Phi)~.~~~\eea
$\mathcal{D}(\Phi)$ is a function of $\Phi$, and in general can be
decomposed into a free part $D_0$ and an interaction part $V$ as
$\mathcal{D}^{\alpha}_{~\beta}(\Phi)=(D_0)^{\alpha}_{~\beta}+V^{\alpha}_{~\beta}(\Phi)$.
The $\mathcal{Z}^{\Lambda}(\Phi)$ can be dropped at tree-level.
After some evaluation including LSZ reduction for fields
$H(p_1+zq)$, $H(p_n-zq)$, the remaining part yields
\bea
\mathcal{Z}(z)=\epsilon^1_{\alpha_1}\epsilon^{n}_{\alpha_n}\Big[V^{\alpha_1\alpha_n}-V^{\alpha_1\beta_1}(D_0^{-1})_{\beta_1\beta_2}V^{\beta_2\alpha_2}+\cdots\Big]~.~~~\eea
Then we can read out the $z$-dependence from above result.

%%%%%%%%%%%%%%%%%%%%%%%%%%
\section{Discussion on the large $z$ behavior}
\label{largeZN4}
%%%%%%%%%%%%%%%%%%%

Let us start from the Lagrangian of $\mathcal{N}=4$ SYM in component
fields,
\bea L&=&-\frac{1}{4}F^a_{\mu\nu}F^{\mu\nu
a}-\frac{1}{2}\mathcal{D}_{\mu}\phi^{Ia} \mathcal{D}^{\mu}\phi^{Ia}
-i\bar{\psi}_A^a\bar{\sigma}^{\mu}\mathcal{D}_{\mu}\psi^{Aa}\nonumber\\
&&+\frac{ig}{2}f^{abc}\left(\bar{T}^I_{AB}\phi^{Ia}\psi^{Ab}\psi^{Bc}
+T^{IBA}\phi^{Ia}\bar{\psi}^b_A\bar{\psi}^c_B\right)
-\frac{g^2}{4}f^{abe}f^{cde}\phi^{Ia}\phi^{Jb}\phi^{Ic}\phi^{Jd}~,~~~\eea
where $T^{IAB}$ is the transformation matrix between $SO(6)$ and
$SU(4)$ representations of scalar fields $\phi^{AB}={1\over
\sqrt{2}}\phi^{I}T^{IAB}$. The gauge fixing term is
\bea  L_{gf}=-\frac{1}{2}(\mathcal{D}_{\mu}A^{\Lambda \mu
a}+gf^{abc}\phi^{Ib}\phi^{\Lambda Ic})^2~.~~~\label{bggf}\eea
In order to get the quadratic terms of shifted hard fields, we need
to compute the second order variation of $L$. Since
\bea &&\frac{\delta L}{\delta A^a_{\mu}}=-\mathcal{D}_{\nu}F^{\mu\nu
a}-gf^{abc}\phi^{Ib} D^{\mu}\phi^{Ic}
-igf^{abc}\bar{\psi}_A^c\bar{\sigma}^{\mu}\psi^{Ab}~,~~~\nonumber\\
&&\frac{\delta L}{\delta \phi^{Ia}}=\mathcal{D}^2\phi^{Ia}
+\frac{ig}{2}f^{abc}\left(\bar{T}^I_{AB}\psi^{Ab}\psi^{Bc}
+T^{IBA}\bar{\psi}^b_A\bar{\psi}^c_B\right)
-g^2f^{abe}f^{cde}\phi^{Jb}\phi^{Ic}\phi^{Jd}~,~~~\nonumber\\
&&\frac{\delta L}{\delta \bar{\psi}_A^a}=
-i\bar{\sigma}^{\mu}D_{\mu}\psi^{Aa}
+igf^{abc}T^{IBA}\phi^{Ic}\bar{\psi}^b_B~,~~~\nonumber\\
&&\frac{\delta L}{\delta \psi^{Aa}}=
i\sigma^{\mu}D_{\mu}\bar{\psi}_A^a
+igf^{abc}\bar{T}^I_{AB}\phi^{Ic}\psi^{Bb}~,~~~\nonumber\eea
we have
\bea D&=&\left(
       \begin{array}{c}
        \frac{\delta }{\delta A^a_{\mu}}   \\ \frac{\delta }{\delta \phi^{Ia}}  \\ \frac{\delta }{\delta \bar{\psi}_A^a}
 \\\frac{\delta }{\delta \psi^{Aa}} \\
       \end{array}
     \right)L\left(
               \begin{array}{cccc}
                 \frac{\overleftarrow{\delta} }{\delta A^b_{\nu}}  & \frac{\overleftarrow{\delta} }{\delta \phi^{Jb}} & \frac{\overleftarrow{\delta} }{\delta \bar{\psi}_B^b} & \frac{\overleftarrow{\delta} }{\delta \psi^{Bb}} \\
               \end{array}
             \right)\nonumber\\
&=&\left(
     \begin{array}{cccc}
      D_{11}&-2gf^{abc}\mathcal{D}^{\mu}\phi^{J c}&igf^{abc}\psi^{Bc}\sigma^{\mu}
&igf^{abc}\bar{\psi}_B^c\bar{\sigma}^{\mu} \\
2gf^{abc}\mathcal{D}^{\nu}\phi^{I c}
&D_{22}
&-igf^{abc}T^{IBA}\bar{\psi}^c_A
&-igf^{abc}\bar{T}^I_{AB}\psi^{Ac}\\
-igf^{abc}\bar{\sigma}^{\nu}\psi^{Ac}
&-igf^{abc}T^{JBA}\bar{\psi}^c_B
&igf^{abc}T^{IBA}\phi^{Ic}
&-i\delta^A_B\bar{\sigma}^{\mu}\mathcal{D}^{ab}_{\mu}\\
igf^{abc}\sigma^{\nu}\bar{\psi}_A^c
&-igf^{abc}\bar{T}^J_{AB}\psi^{Bc}
&i\delta_A^B\sigma^{\mu}\mathcal{D}^{ab}_{\mu}
&igf^{abc}\bar{T}^I_{AB}\phi^{Ic}\\
     \end{array}
   \right)~,~~~
  \eea
where
\bea &&D_{11}=\eta^{\mu\nu}\left[(\mathcal{D}^2)^{ab}-g^2f^{ace}f^{bde}\phi^{Kc}\phi^{Kd}\right]-2gf^{abc}F^{\mu\nu c}~,~~~\\
&&D_{22}=\delta^{IJ}\left[(\mathcal{D}^2)^{ab}-g^2f^{ace}f^{bde}\phi^{Kc}\phi^{Kd}\right]
-2g^2f^{abc}f^{cde}\phi^{Id}\phi^{Je}~,~~~\eea
and $\mathcal{D}^{-ab}=\delta^{ab}\partial^--gf^{abc}A^{-c}$. The
operator $D$ can be decomposed into two parts, the interaction part
$V(z)=V+zX$, where
\bea X=\left(
         \begin{array}{cccc}
    2igf^{abc}\eta^{\mu\nu}A^{-c}
&2igf^{abc}q^{\mu}\phi^{J c}
&0&0 \\
-2igf^{abc}q^{\nu}\phi^{I c}
&2igf^{abc}\delta^{IJ}A^{-c}
&0&0\\
0&0&0
&-\delta^{ab}\delta^A_Bq_{\mu}\bar{\sigma}^{\mu}\\
0&0
&\delta^{ab}\delta_A^Bq_{\mu}\sigma^{\mu}
&0  \\
         \end{array}
       \right)~,~~~
\eea
and the free field part
\bea D_0=\delta^{ab}\left(
                      \begin{array}{cccc}
                      \eta^{\mu\nu}\partial^2&0&0&0\\
0&\delta^{IJ}\partial^2&0&0\\
0&0&0
&-i\delta^A_B\bar{\sigma}^{\mu}\partial_{\mu}\\
0&0&i\delta_A^B\sigma^{\mu}\partial_{\mu}
&0\\
                      \end{array}
                    \right)~,~~~
\eea
where we can write
$D_0^{-1}=d_0+\frac{d_1}{z}+\frac{d_2}{z^2}+\mathcal{O}(\frac{1}{z^3})$.
Defining
\bea \delta_0=\left(
           \begin{array}{cccc}
            \eta_{\mu\nu}&0&0&0\\
0&\delta^{IJ}&0&0\\
0&0&0
&-i\delta_A^B\bar{\sigma}^{\mu}\partial_{\mu}\\
0&0&i\delta^A_B\sigma^{\mu}\partial_{\mu}
&0\\
           \end{array}
         \right)~~~,~~~\delta_1=\left(
                                  \begin{array}{cccc}
                                 0&0&0&0\\
0&0&0&0\\
0&0&0
&-\delta_A^B\bar{\sigma}^{\mu}q_{\mu}\\
0&0&\delta^A_B\sigma^{\mu}q_{\mu}
&0\\
                                  \end{array}
                                \right)~,~~~\eea
we have
\bea &&D_0^{-1}=\delta^{ab}(\partial^2)^{-1}\delta_0~~,~~
d_0=\delta^{ab}\frac{i\delta_1}{2\partial^-}~,~~~\nonumber\\
&&d_1=\delta^{ab}[\frac{\partial^2\delta_1}{4(\partial^-)^2}+\frac{i\delta_0}{2\partial^-}]~~,~~
d_2=\delta^{ab}[-\frac{i(\partial^2)^2\delta_1}{8(\partial^-)^3}
+\frac{\partial^2\delta_0}{4(\partial^-)^2}]~.~~~\eea
It is very crucial to have $d_0X=Xd_0=0$, then the expansion
\bea
V(z)\big(1+D_0^{-1}V(z)\big)^{-1}&=&(V+zX)\left(1+(d_0+\frac{d_1}{z}+\frac{d_2}{z^2}+\cdots)(V+zX)\right)^{-1}\nonumber\\
&=&zX(1+d_1X)^{-1}+\mathcal{O}(z^0)~.~~~\eea

Now let us first consider $\spab{g_1^-|g_n^+}$-shifting, and
determine the leading order of
$\mathcal{Z}^{\spab{g_1^-|g_n^+}}(z)$. The helicity vectors of
$g_{\widehat{1}}^-,g_{\widehat{n}}^+$ both introduce a factor of
$z$, while $zX(1+d_1X)^{-1}$ introduce another factor of $z$. Notice
that both $d_1$ and $X$ are block-diagonal, which means fermion
operators will not appear. It implies that, in this order, the
$\mathcal{Z}^{\spab{g_1^-|g_n^+}}(z)$ of $\mathcal{N}=4$ SYM is the
same as its bosonic sub-theory, which is a 4-dimensional reduction
of 10-dimensional Yang-Mills theory. According to
\cite{Jin:2015pua},
\bea
\mathcal{Z}^{\spab{g_1^{-a}|g_n^{+b}}}(z)=-2iz^3gf^{abc}(p_1\cdot
p_n)q_{\mu}A^{\mu c}+O(z^2)~.~~~\eea
The leading order is $z^3$. If the color indices of two shifted
fields are contracted, then the first term vanishes due to
$f^{abc}=0$ when $a=b$, and the leading order becomes $z^2$, while
we know in \S \ref{sectionspin2} that the leading order of double
trace term under such shifting is $z^4$.

In paper \cite{Elvang:2008na}, it was proved that the large $z$
behavior of $\Spab{\Phi^{U_1a}|\Phi^{U_2b}}$-shifting is
\bea
\mathcal{Z}^{\Spab{\Phi^{U_1a}|\Phi^{U_2b}}}(z)=O(z^{|U_1/U_n|-1})~.~~~\eea
We would like to refine their result as
\bea
\mathcal{Z}^{\Spab{\Phi^{U_1a}|\Phi^{U_2b}}}(z)=z^{|U_1/U_n|-1}f^{abc}\mathcal{L}^c_{\spab{\Phi^{U_1a}|\Phi^{U_2b}}}+O(z^{|U_1/U_n|-2})~,~~~\label{Zrefine}\eea
where $\mathcal{L}^c$ is an arbitrary operator, and there is always
a $f^{abc}$ associated with the leading order term. We already
proved (\ref{Zrefine}) for $\langle g^{-a}|g^{+b}]$-shifting. Since
all states in $\mathcal{N}=4$ SYM are related by SUSY,
(\ref{Zrefine}) also holds for any shifting, and the proof will be
complete parallel to \S 7.1 of \cite{Elvang:2008na}. This means that
after contracting the indices $a,b$, the first term vanishes, and
the large $z$ behaves even better than expected.

%%%%%%%%%%%%%

\bibliographystyle{JHEP}
\bibliography{Form}

%\begin{thebibliography}{References}
%\end{thebibliography}

\end{document}